\documentclass[aps,prl,twocolumn,longbibliography,superscriptaddress]{revtex4-2}
\usepackage{bbm}
\usepackage{graphicx}
\usepackage{dcolumn}
\usepackage{bm}
\usepackage{subfigure}
\usepackage{amsmath}
\usepackage{feynmf}
\usepackage{hyperref}
\usepackage{CJK}
\usepackage{amssymb}
\usepackage{attachfile}
\usepackage{multirow}
\usepackage{makecell}
\newcommand{\bk}{\boldsymbol k}
\newcommand{\bp}{\boldsymbol p}

\newcommand{\bq}{\boldsymbol q}

\newcommand{\bd}{\boldsymbol d}

\newcommand{\bOmega}{\boldsymbol{\Omega}}
\newcommand{\bS}{\boldsymbol{S}}
\newcommand{\bsigma}{\boldsymbol{\sigma}}

\newcommand{\boldm}{\boldsymbol{m}}

\newcommand{\bl}{\boldsymbol{l}}

\newcommand{\zb}{\color {black}}
\newcommand{\zy}{\color {black}}

\usepackage{braket}
\usepackage{esint}
\usepackage{times}

\begin{document}

\title{Cartesian Nodal Lines and Magnetic Kramers Weyl Nodes in Spin-Split Antiferromagnets}

\author{Zheng-Yang Zhuang}
\affiliation{Guangdong Provincial Key Laboratory of Magnetoelectric Physics and Devices,
	State Key Laboratory of Optoelectronic Materials and Technologies,
	School of Physics, Sun Yat-sen University, Guangzhou 510275, China}

\author{Di Zhu}
\affiliation{Guangdong Provincial Key Laboratory of Magnetoelectric Physics and Devices,
	State Key Laboratory of Optoelectronic Materials and Technologies,
	School of Physics, Sun Yat-sen University, Guangzhou 510275, China}

\author{Zhigang Wu}
\affiliation{Quantum Science Center of Guangdong-Hong Kong-Macao Greater Bay Area (Guangdong), 
	Shenzhen 508045, China}

\author{Zhongbo Yan}
\email{yanzhb5@mail.sysu.edu.cn}
\affiliation{Guangdong Provincial Key Laboratory of Magnetoelectric Physics and Devices,
	State Key Laboratory of Optoelectronic Materials and Technologies,
	School of Physics, Sun Yat-sen University, Guangzhou 510275, China}

\date{\today}

\begin{abstract}
	When band degeneracy occurs in a spin-split band structure, it gives rise to divergent Berry curvature and distinctive topological boundary states, resulting in a variety of fascinating effects. We show that three-dimensional spin-split antiferromagnets, characterized by symmetry-constrained momentum-dependent spin splitting and zero net magnetization, can host two unique forms of symmetry-protected band degeneracy: Cartesian nodal lines in the absence of spin-orbit coupling, and magnetic Kramers Weyl nodes when spin-orbit coupling is present. Remarkably, these band degeneracies not only produce unique patterns of Berry-curvature distributions but also give rise to topological boundary states with unconventional spin textures. Furthermore, we find that these band degeneracies can lead to strong or even  quantized anomalous Hall effects and 
	quantized circular photogalvanic effects under appropriate conditions. Our study suggests that spin-split antiferromagnets provide a fertile ground for exploring unconventional topological phases.  
\end{abstract}

\maketitle

{\it Introduction.---}Momentum-dependent spin splitting (MDSS) serves as a pivotal driving force behind the emergence of non-trivial quantum geometry and the realization of a diverse range of topological phases. Spin-orbit coupling (SOC) in 
noncentrosymmetric systems is a fundamental mechanism responsible for this phenomenon and 
has sparked extensive research over the past two decades~\cite{Galitski2013,Sinova2015,Manchon2015,Schaffer2016,Soumyanarayanan2016,Smidman2017}. 
Due to its time-reversal ($\mathcal{T}$)-even and inversion ($\mathcal{P}$)-odd nature, SOC-induced MDSS exhibits intrinsic 
nodes at time-reversal-invariant momentums (TRIMs). These nodes manifest as
band degeneracies within the band structure and underpin various topological phases~\cite{Fu2008TSC,Sato2009TSC,Lutchyn2010,Oreg2010,Yu2010QAHE,Chang2013QAHE}.

Exchange interaction is another fundamental mechanism  for spin splitting. Recent discoveries have revealed that even in antiferromagnets
with zero-net magnetization, spin splitting can exhibit significant strength and momentum 
dependence~\cite{Hayami2019AM,Hayami2020AM,Yuan2020AM,Yuan2021AM,
	Liu2022AMPRX,Ma2021AM,Shao2021NC,Libor2022AMa,Libor2022AMb,Libor2022AMc,Zhu2024,Liu2024AMPRX,Xiao2024SSG,Jiang2024SSG}, provided that the system lacks 
$\mathcal{PT}$ symmetry or $\mathcal{T}\tau$ symmetry, where $\tau$ represents a translation operation. 
Notably,  the exchange-interaction-induced MDSS in spin-split antiferromagnets also features nodes, 
resulting in symmetry-enforced band degeneracies. For example, this is seen in 
spin-split antiferromagnets with collinear magnetic moments, also known as altermagnets~\cite{Libor2022AMb,Mazin2022AM}. These materials have garnered significant attention due to their unique spin-split band structures~\cite{Osumi2024MnTe,Lee2024MnTe,Krempasky2024,Hajlaoui2024AM,Reimers2024,Ding2024CrSb,Yang2025CrSb,
	Zeng2024CrSb,Li2025CrSb,Lu2025CrSb,Jiang2025} and a wide range of intriguing phenomena they host~\cite{Ouassou2023AM,
	Lu2024AM11,Zhang2024AM,Sumita2023FFLO,Chakraborty2024AM8,Zhu2023TSC,Zhu2024dislocation,Ghorashi2024AM,Li2024AMHOTI,Rao2024AM7,Li2025AM1,
	Brekke2023AM,Kristian2024AM,Wei2024AM,Carvalho2024AM,
	Attias2024AM9,Fang2023NHE,Jin2024AM11,Hu2025NLME,Hu2025AM1,
	Antonenko2025AM,Parshukov2025AM,Tan2024AM7,Heung2025AM,Zhang2024AM12,Jiang2025AM,Hoyer2025AM,Wei2024AM9,
	Hu2024AM10,Ezawa2025AM12,Chen2025AM1,Duan2025AFMAM,Samanta2025AM8,Chen2025AM12,Lin2025AM12,Qu2025,Zhu2025}. 
As spin is conserved for a collinear magnetic order,  the nodes in the MDSS
form nodal surfaces in three dimensions (3D) and nodal lines in two dimensions (2D)
in the absence of SOC. However, these band degeneracies are characterized by a codimension of $d_{c}=1$ 
and generally do not lead to topological boundary states~\cite{Wu2018NSS}. 

In search of band degeneracies with nontrivial properties,   
we are led to examine spin-split antiferromagnets with noncollinear magnetic moments~\cite{Hayami2020AM,Yuan2021AM,Zhu2024}. 
For these materials, the spin conservation is intrinsically broken and 
the nodes in the exchange-interaction-induced MDSS are severely constrained by crystal symmetries. 
By analyzing additional constraints imposed by these symmetries and the interplay of SOC and exchange interaction, we uncover two unique forms of band degeneracies with fascinating properties, which we refer to as the {\it Cartesian nodal lines} (CNLs) and {\it magnetic Kramers Weyl nodes} (MKWNs) respectively.  

Without SOC, band degeneracies in 3D spin-split noncollinear antiferromagnets generally take the  form of 
nodal lines with a codimension $d_{c}=2$~\cite{Fernandes2023}. 
Since the $\mathcal{PT}$ symmetry is absent, these nodal lines are protected by mirror symmetry and 
are confined to mirror planes. Additional crystal symmetries further constrain them to intersect and form a structure resembling the 
Cartesian coordinate system, hence the name {\it Cartesian nodal lines}.  Distinguished from other nodal-line structures 
protected by $\mathcal{PT}$ symmetry or chiral symmetry~\cite{Burkov2011nlsm,Kim2015,Yu2015,Fang2015nodal,
	Chen2017link,Yan2017link,Ren2017knot,Ezawa2017link,Chang2017link,Ikegaya2017},
these CNLs give rise to not only unique Berry curvature distributions but also topological 
surface states with unconventional spin textures. In the presence of SOC, these CNLs undergo a transition into Weyl 
nodes with $d_{c}=3$. Notably,  some of these Weyl nodes are pinned at specific TRIMs. Reminiscent of the 
Kramers Weyl nodes protected by time-reversal symmetry in chiral crystals~\cite{Chang2018,Hasan2021},  these Weyl nodes are thus dubbed {\it magnetic Kramers Weyl nodes}. The existence of these band degeneracies has far-reaching consequences, as they can lead to, under suitable conditions, strong or even quantized anomalous Hall effects as well as quantized circular
photogalvanic effects.

\begin{figure}[t]
	\centering
	\subfigure{
		\includegraphics[scale=0.35]{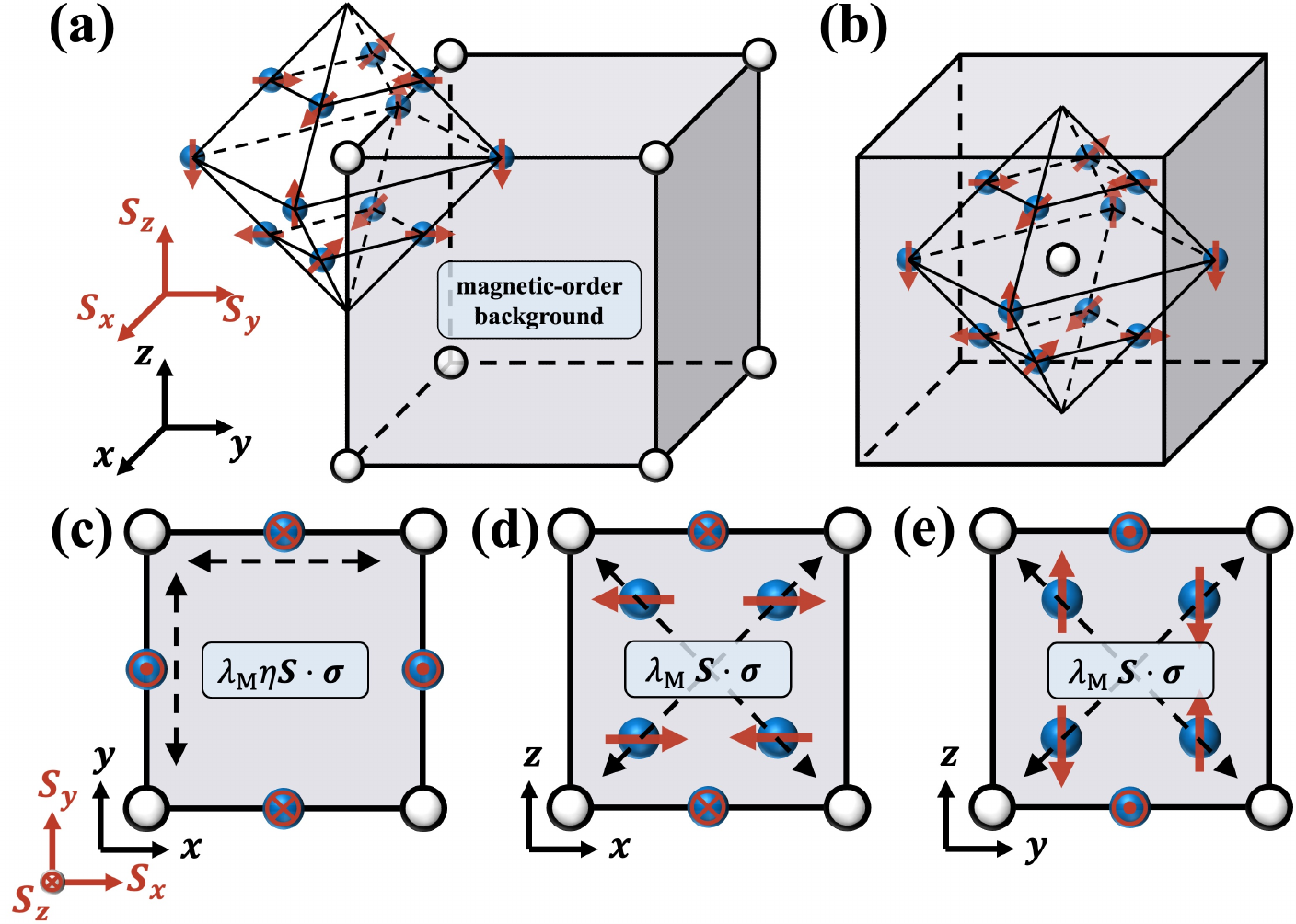}}
	\caption{Schematic of the lattice structure. White (blue) spheres represent nonmagnetic (magnetic) atoms, while red arrows indicate the local magnetic moments. [(a)] presents the magnetic background surrounding a representative nonmagnetic atom; the configurations around other nonmagnetic atoms are identical and thus omitted for clarity. (b) Schematic illustration of the Wigner-Seitz unit cell. The spin (red) and real-space (black) coordinate conventions are the same as in [(a)]. (c-e) Schematics of the spin-dependent hoppings of the form $\bS\cdot\bsigma$ in the $xy$, $xz$, and $yz$ planes, respectively. Dashed black arrows indicate hopping directions, and the corresponding amplitudes are shown in light-blue rounded rectangles. Panels [(d,e)] share the same spin-coordinate convention as [(c)].}
	\label{Fig1}
\end{figure}

{\it Cartesian nodal lines.---}Constrained by symmetry, the exchange-interaction-induced MDSS can be viewed as 
an order parameter analogous to the superconducting pairing~\cite{Jungwirth2025review}, and can be classified by the irreducible representations of 
symmetry groups~\cite{Fernandes2023}. While the physics discussed here is general, we focus on a tetragonal-lattice 
antiferromagnet belonging to the $D_{2h}$ magnetic point group of the $D_{4h}$ point group (magnetic group $D_{4h}$ ($D_{2h}$) in the $G(H)$ convention).
A schematic of the lattice structure is shown in Fig.~\ref{Fig1}. We assume that the electrons associated with magnetic atoms 
are strongly localized, and the itinerant electrons, which are responsible for the system's transport behaviors, 
originate from nonmagnetic atoms. The magnetic atoms act as a magnetic background and influence
the itinerant electrons as spin-dependent potentials.  
When the itinerant electrons hop along paths intersecting these magnetic atoms, 
their hoppings become spin-dependent as an electron with spin parallel or antiparallel to the 
local magnetic moment experiences a different tunneling probability. As a result,  
the tight-binding Hamiltonian describing  
the spin-split band structure of itinerant electrons takes the form $\hat{H}=\sum_{k}c_{\bk}^{\dagger}\mathcal{H}(\bk)c_{\bk}$, where $c_{\bk}^{\dagger}=(c_{\bk\uparrow}^{\dagger},c_{\bk\downarrow}^{\dagger})$ and 
\begin{equation}
	\mathcal{H}(\bk)=\varepsilon_{0}(\bk)\sigma_{0}+\lambda_{\rm so}\bl(\bk)\cdot\bsigma+\lambda_{\rm M}\boldm(\bk)\cdot\bsigma.
	\label{eq: H}
\end{equation}
Here, $\bsigma=(\sigma_{x},\sigma_{y},\sigma_{z})$ denotes the vector of Pauli matrices, and $\sigma_{0}$
represents the identity matrix.
The first term, with $\varepsilon_{0}(\bk)=-t(\cos k_{x}+\cos k_{y})-t_{z}\cos k_{z}$ refers to the kinetic energy. The second term represents spin-orbit coupling (SOC), $\bl(\bk)=(\sin k_{x},\sin k_{y}, \sin k_{z})$, which is forbidden by the spatial symmetries (including inversion and mirror symmetries) of the lattice in Fig.~\ref{Fig1}, but can emerge once these symmetries are broken, for instance by lattice distortion. The last term accounts for the magnetic exchange field~\cite{Fernandes2023} with $\boldm(\bk)=[-\sin k_{x}\sin k_{z}, \sin k_{y}\sin k_{z}, \eta(\cos k_{x}-\cos k_{y})]$, where the parameter $\eta$ characterizes the magnetic anisotropy and is typically nonzero.
For notational simplicity, we set the lattice constants to unity throughout this work.

\begin{figure}[t]
	\centering
	\subfigure{
		\includegraphics[scale=0.35]{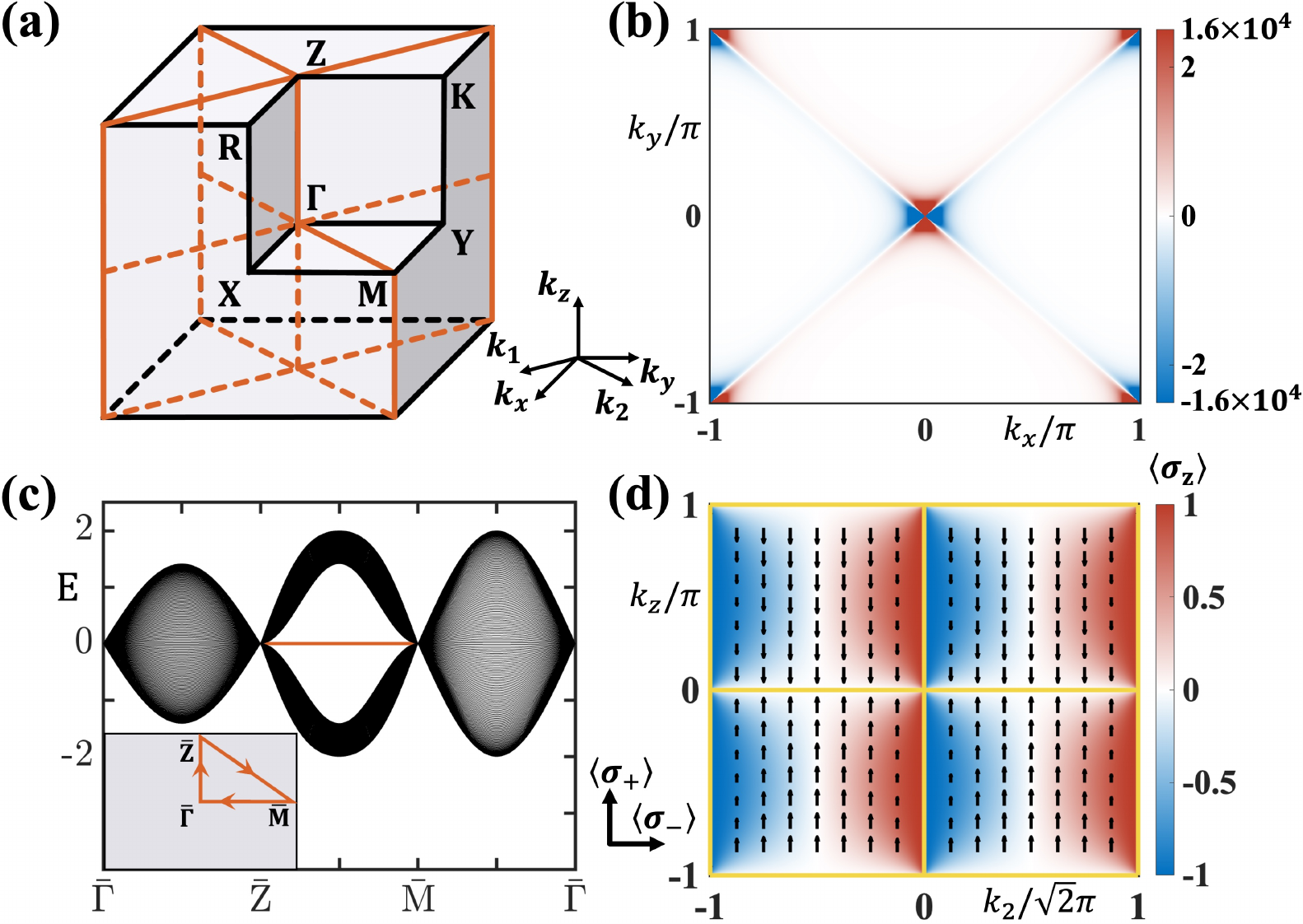}}
	\caption{Cartesian nodal lines and their associated topological properties. (a) Solid and dashed orange lines represent nodal lines, 
		which form a structure analogous to the Cartesian coordinate system at each intersection. 
		(b) Distribution of the Berry curvature in the momentum plane with $k_{z}=0.1$. (c) Energy spectra along a path in the surface Brillouin zone, with open boundary conditions in the $[\bar{1}10]$ direction. High-symmetry points in the surface Brillouin zone are shown in the inset. (d) 
		Black arrows and gradient colors jointly depict the spin polarizations of the surface states on the right $(\bar{1}10)$ surface. 
		The yellow network marks the boundary of the zero-energy surface states.  Common parameters are $t=t_{z}=\lambda_{\rm so}=0$, and $\lambda_{\rm M}=\eta=1$.}
	\label{Fig2}
\end{figure}

We begin by examining the MDSS arising purely from the magnetic exchange field in the lattice shown in Fig.~\ref{Fig1}, where SOC is forbidden.
In this case, the system respects inversion symmetry, 
three mirror symmetries ($\mathcal{M}_{z}$, $\mathcal{M}_{xy}$, $\mathcal{M}_{\bar{x}y}$), and
combined rotation-time-reversal symmetries ($C_{4z}\mathcal{T}$, $C_{2x}\mathcal{T}$, 
$C_{2y}\mathcal{T}$), where $C_{na}$ represents a
$2\pi/n$ rotation about the $a$ axis. These $C_{na}\mathcal{T}$ symmetries 
ensure a zero net magnetization. The corresponding energy spectra are $E_{\pm}(\bk)=\varepsilon_{0}(\bk)\pm\lambda_{\rm M}\sqrt{\sin^{2}k_{z}(\sin^{2}k_{x}+\sin^{2}k_{y})+\eta^{2}(\cos k_{x}-\cos k_{y})^{2}}$.
It is clear from the spectra that band degeneracies 
appear at these positions: (1) along the momentum lines satisfying $|k_{x}|=|k_{y}|$ within the $k_{z}=0/\pi$ planes, and (2) along the high-symmetry $k_{z}$ lines passing through $(k_{x},k_{y})=(0/\pi, 0/\pi)$, as illustrated in Fig.~\ref{Fig2}(a).
Owing to the polar-vector nature of spin---spin components parallel to a mirror plane flip sign, while those perpendicular remain invariant---these nodal lines are protected by the three mirror symmetries. They intersect orthogonally 
at the four $C_{4z}\mathcal{T}$-invariant momenta within the Brillouin zone, i.e., $(0,0,0/\pi)$
and $(\pi,\pi,0/\pi)$. At each intersection, 
the nodal-line structure is analogous to the Cartesian coordinate system, motivating the term CNL. 
While CNLs can be considered a distinct class of crossed 
$Z_{3}$ nodal nets, their origin and properties differ significantly from those found in nonmagnetic~\cite{S.A.Yang2017PRB} and altermagnetic materials~\cite{XZhang2024PRL}.

Unlike nodal-line semimetals protected by $\mathcal{PT}$ symmetry~\cite{Fang2016review}, where the Berry curvature is identically zero~\cite{Xiao2010review}, mirror-protected CNLs permit finite Berry curvature, which
becomes divergently 
large near the band degeneracy, as illustrated in Fig.~\ref{Fig2}(b). The distribution of the Berry curvature 
respects the symmetries of the point group, resulting in an exact cancellation of the Hall conductivity when the anomalous Hall 
effect is considered~\cite{Nagaosa2010RMP}.  Nevertheless, its divergent nature near the band degeneracy implies that 
an external perturbation, which disrupts the symmetry-enforced cancellation, can 
induce a strong anomalous Hall effect. This will be demonstrated in detail later. 

A characteristic of nodal lines is the emergence of dispersionless topological surface states when 
chiral symmetry is present~\cite{Schnyder2011flat}. These topological surface states exhibit fixed spin polarizations, 
as they simultaneously serve as eigenstates of the chiral symmetry operator—a constant unitary 
and Hermitian operator that anticommutes with the Hamiltonian~\cite{Schnyder2008}. 
Intriguingly, we discover that in this system, topological surface flat bands 
exist even in the absence of chiral symmetry, provided that the term $\varepsilon_{0}(\bk)\sigma_{0}$, 
which has no impact on topology, is omitted from the Hamiltonian. In Fig.~\ref{Fig2}(c), 
the energy spectrum for a system with open boundary conditions along the $[\bar{1}10]$ direction is plotted 
along a specific high-symmetry path in the surface Brillouin zone. 
The existence of surface flat bands at zero energy is clearly visible. Furthermore, the region supporting topological 
surface states spans the whole surface Brillouin zone, except for these four high-symmetry lines
where the bulk spectrum is gapless. This is because the CNLs form a network and their projection 
along the $[\bar{1}10]$ direction overlaps with the four high-symmetry lines of the surface Brillouin zone.  
By analyzing the spin textures of these topological 
surface states, we find that their spin polarizations are not fixed but instead 
exhibit strong momentum dependence, as illustrated in Fig.~\ref{Fig2}(d). Interestingly, 
the spin textures in the four quadrants of the surface Brillouin zone are symmetry-related 
and form a plaid-like configuration. This distinctive pattern of spin textures can 
serve as a definitive signature for identifying the presence of CNLs in experiments.  

To understand the existence of topological surface flat bands in the absence of 
chiral symmetry and the accompanying spin textures, we invoke the concept of subchiral symmetry, as introduced in Ref.~\cite{Mo2024subchiral}. Subchiral symmetry plays the role of chiral symmetry along one-dimensional momentum lines when the perpendicular momentum components are treated as parameters, even though global chiral symmetry is broken. This generalization of chiral symmetry has proven useful in various systems~\cite{Liu2024Dirac,Biao2024subchiral}.
To make the subchiral symmetry explicit, we rotate the coordinate frame by $\pi/4$ around the $z$-axis. In the rotated frame, the momentum ${\boldsymbol k} = (k_1,k_2,k_z)$, where $k_{1}=\frac{k_{x}-k_{y}}{\sqrt{2}}$, $k_{2}=\frac{k_{x}+k_{y}}{\sqrt{2}}$ [see Fig.~\ref{Fig2}(a)], and the Pauli-matrix vector transforms to $\bsigma=(\sigma_{-},\sigma_{+},\sigma_{z})$, where $\sigma_{\pm}=\frac{\sigma_{x}\pm\sigma_{y}}{\sqrt{2}}$. 
The SOC-free Hamiltonian, denoted by $\mathcal{H}_{\rm CNL}(\bk)$, becomes (see Supplemental information (SI) Sec. I.A for details)
\begin{eqnarray}
	\mathcal{H}_{\rm CNL}(\bk)&=&-\sqrt{2}\lambda_{\rm M}\sin k_{z}\cos\frac{k_{1}}{\sqrt{2}}\sin\frac{k_{2}}{\sqrt{2}}\ \sigma_{-}\nonumber\\
	&&-\lambda_{\rm M}\Lambda(k_{2},k_{z})\sin\frac{k_{1}}{\sqrt{2}}\sigma_{g}(k_{2},k_{z}),
	\label{eq: transed H}
\end{eqnarray}
where $\sigma_{g}(k_{2},k_{z})=\sin\theta(k_{2},k_{z})\sigma_{z}+\cos\theta(k_{2},k_{z})\sigma_{+}$, $\theta(k_{2},k_{z})=\arg(\sqrt{2}\sin k_{z}\cos\frac{k_{2}}{\sqrt{2}}+i 2\eta \sin \frac{k_{2}}{\sqrt{2}})$
and $\Lambda(k_{2},k_{z})=\sqrt{2\sin^{2} k_{z}\cos^{2}\frac{k_{2}}{\sqrt{2}}+4\eta^{2} \sin^{2} \frac{k_{2}}{\sqrt{2}}}$. 
We have dropped the term $\varepsilon_{0}(\bk)\sigma_{0}$ as it does not affect the band topology. 
Crucially, $\sigma_{g}(k_{2},k_{z})$ is unitary and anticommutes with $\sigma_{-}$. Hence, there exists a third unitary matrix $\mathcal{S}(k_{2},k_{z})=-\sin\theta(k_{2},k_{z})\sigma_{+}+\cos\theta(k_{2},k_{z})\sigma_{z}$ that anticommutes with both $\sigma_{g}(k_{2},k_{z})$ and $\sigma_{-}$, completing the Pauli-matrices-like Clifford algebra. This matrix $\mathcal{S}(k_{2},k_{z})$ defines the subchiral symmetry operator
since $\{\mathcal{S}(k_{2},k_{z}),\mathcal{H}_{\rm CNL}(\bk)\}=0$.
For surface states on the $(\bar{1}10)$ surfaces, 
$k_{2}$ and $k_{z}$ are good quantum numbers and hence can be viewed as parameters. 
For fixed $k_{2}$ and $k_{z}$, $\mathcal{S}(k_{2},k_{z})$ becomes a constant matrix similar to the operator for chiral symmetry, and accordingly, subchiral symmetry plays the role of chiral symmetry. Therefore, the zero-energy topological surface states presented in 
Fig.~\ref{Fig2}(c) can be characterized by the following winding number~\cite{Ryu2010},
$W(k_{2},k_{z})
=\frac{1}{4\pi i}\int_{-\sqrt{2}\pi}^{\sqrt{2}\pi}dk_{1}{\rm Tr}\left[\mathcal{S}(\mathcal{H}_{\rm CNL})^{-1}\partial_{k_{1}}\mathcal{H}_{\rm CNL}\right]$.
The spectrum along the $k_{1}$ direction exhibits insulating behavior as long as $k_{2}\neq\{0,\sqrt{2}\pi\}$ and $k_{z}\neq\{0,\pi\}$. 
In the gapped region, we find $W(k_{2},k_{z})=\text{sgn}(k_{2}k_{z})$, where 
$\text{sgn}(\cdot)$ denotes the sign function. The nonzero winding number indicates the presence of zero-energy states at the boundary, which collectively form the surface flat bands. Since these zero-energy surface states simultaneously serve as eigenstates of 
the subchiral symmetry operator $\mathcal{S}(k_{2},k_{z})$. Their spin textures follow directly:~(see Note S1 in SI)
\begin{equation}
	\begin{aligned}
		\braket{\sigma_{z}}(k_{2},k_{z})&=\beta {\rm sgn}(k_{2}k_{z})\cos\theta(k_{2},k_{z}),\\
		\braket{\sigma_{+}}(k_{2},k_{z})&=-\beta {\rm sgn}(k_{2}k_{z})\sin\theta(k_{2},k_{z}),\\
		\braket{\sigma_{-}}(k_{2},k_{z})&=0,
	\end{aligned}
\end{equation}
where $\beta=1$ $(-1)$ refers to the left (right) surface, indicating that the spin polarization on opposing 
surfaces is reversed. These analytical results provide an intuitive and consistent explanation for
the numerical observations in Figs.~\ref{Fig2}(c) and (d).

\begin{figure}[t]
	\centering
	\subfigure{
		\includegraphics[scale=0.35]{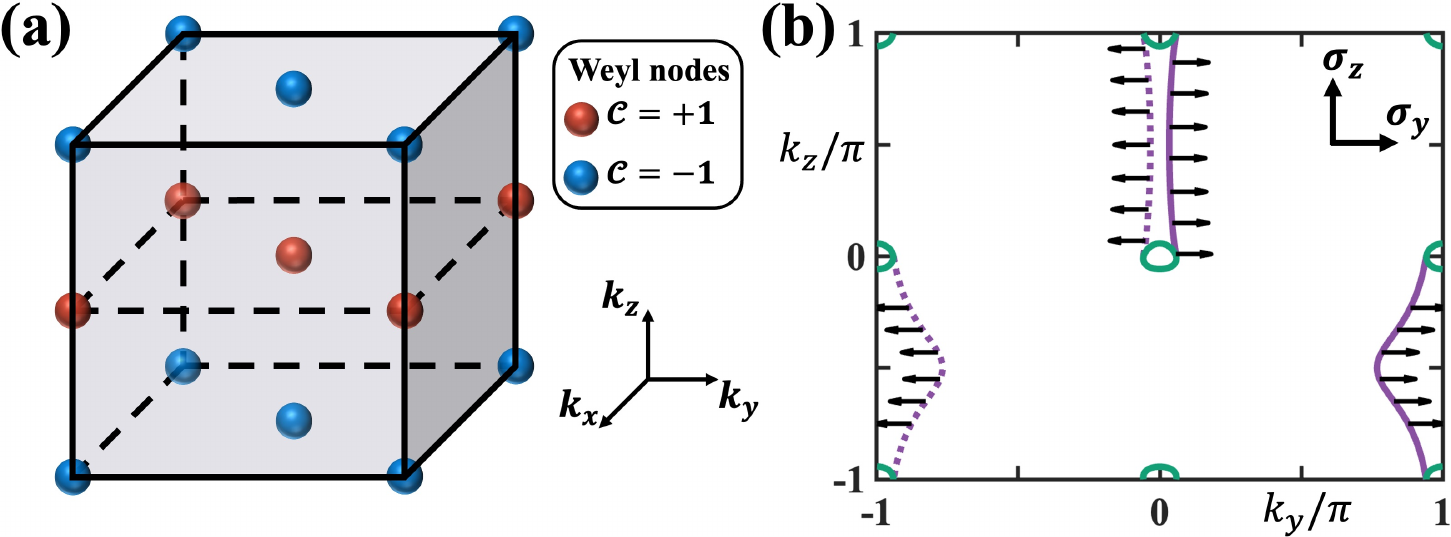}}
	\caption{Magnetic Kramers Weyl nodes and their associated topological surface states. (a) Schematic of MKWNs in Brillouin zone when $\eta>\eta_{c}$. Blue and red spheres represent MKWNs with opposite charges, illustrated in the middle inset. (b) Fermi arcs on the left (right) $x$-normal surface is denoted by solid (dotted) purple lines, and the black arrows denote their spin polarizations. The solid green rings represent the projections of bulk states at fixed energy $E=0.2$. Parameters are $t=t_{z}=0$, $\lambda_{\rm M}=0.4$, $\lambda_{\rm so}=1.1$, and $\eta=1$.}
	\label{Fig3}
\end{figure}

{\it Magnetic Kramers Weyl nodes.---}Finite SOC is introduced when inversion and mirror symmetries are broken by symmetry-lowering perturbations, such as strain-induced lattice distortion that gives rise to an asymmetric distribution of the electric potential around the nonmagnetic atoms. In this situation, the magnetic point group is reduced from $D_{4h}$ ($D_{2h}$) to $C_{4}$ ($C_{2}$). Although the system becomes inversion-asymmetric, the essential spin physics can still be accurately captured by the two-band Hamiltonian~\cite{Chang2018} in Eq.~(\ref{eq: H}).
The breaking of these mirror symmetries gaps the previously protected CNLs. However, the surviving $C_{4z}\mathcal{T}$ symmetry mandates the existence of Weyl nodes at the four $C_{4z}\mathcal{T}$-invariant momenta, which coincide with the TRIMs.
Since these Weyl nodes are pinned at TRIMs---similar to the 
Kramers Weyl nodes protected by time-reversal symmetry in chiral crystals\cite{Chang2018,Hasan2021}---we refer to 
them as MKWNs to highlight the absence of time-reversal symmetry in this system.
Their topological charges $\mathcal{C}_{\boldsymbol q}$ are determined by 
the Chern number $\mathcal{C}=\frac{1}{2\pi}\oint\bOmega\cdot d\bS$, where $\bOmega$ 
represents the Berry curvature, and the integral is performed over a closed surface 
enclosing the Weyl node located at momentum $\boldsymbol q$~\cite{Armitage2018RMP}. These 
topological charges are illustrated in the inset located 
between Figs.~\ref{Fig3}(a) and (b). 
We note that magnets can also host a distinct class of Weyl fermions pinned at TRIMs, namely Heesch Weyl fermions~\cite{Gao2023Heesch}, which are not protected by an antiunitary symmetry. By contrast, MKWNs require such a symmetry (e.g., $C_{4z}\mathcal{T}$ here) and therefore exhibit a Kramers-like character.

Beyond the symmetry-enforced MKWNs, the interplay between the SOC and the magnetic exchange field gives rise to additional Weyl nodes. 
Consider the case $\lambda_{\rm so}>\lambda_{\rm M}$. When the anisotropic parameter falls below the critical value $\eta_{c}=\lambda_{\rm so}/(2\lambda_{\rm M})$, two additional pairs of Weyl nodes emerge along the high-symmetry $k_{z}$ lines that traverse $(k_{x},k_{y})=(0,\pi)$ or $(\pi,0)$
(see Note S1 in the SI). These topological charges along with their positions (shown by the subscript) are given by $\mathcal{C}_{(0,\pi,-\pi+k_{0})}=\mathcal{C}_{(\pi,0,\pi-k_{0})}=1$ and $\mathcal{C}_{(0,\pi,-k_{0})}=\mathcal{C}_{(\pi,0,k_{0})}=-1$,
where $k_{0}=\arcsin(\eta/\eta_{c})$. As $\eta$ increases, each pair of oppositely charged Weyl nodes on the same high-symmetry $k_{z}$ line, approaches, meets, and eventually annihilates when $\eta$ exceeds the critical value. 
In the regime $\lambda_{\rm so}<\lambda_{\rm M}$, a richer pattern of
additional Weyl nodes emerges; details are provided in the Note S1 in the SI.
It is worth noting that the low-energy linearly dispersive Weyl cones associated with the 
MKWNs exhibit up-down symmetry (type-I), while the Weyl cones emerging at generic 
positions tend to be tilted and typically belong to the type-II class~\cite{soluyanov2015type} if the parameter $\lambda_{\rm so}$ is
much smaller than the hopping parameter $t_{z}$.  

To further characterize these Weyl nodes, we turn to the Fermi arcs which 
depict the iso-energy contours of surface states, connecting the surface projections of Weyl nodes with opposite 
topological charges~\cite{wan2011,xu2011,Weng2015weyl,huang2015weylb}. They serve as a distinctive hallmark indicating the presence of Weyl nodes~\cite{Lv2015weyl,Xu2015weyl}.  In this system, 
the $C_{4z}\mathcal{T}$ symmetry imparts unique features to both the Fermi arcs 
and their associated spin textures. In particular, when only MKWNs exist, 
each surface hosts two distinct Fermi arcs, each spanning across half of the surface Brillouin zone, 
as depicted in Fig.~\ref{Fig3}(b). These arcs are fully spin polarized, 
and  spin 
textures on $x$- and $y$-normal surfaces are related by $C_{4z}\mathcal{T}$ symmetry: the spin polarizations 
align along the $y$ ($x$) direction on $x$ ($y$)-normal surfaces.
When additional Weyl modes are introduced, the connectivity of the Fermi arcs becomes more intricate, yet the spin texture pattern dictated by $C_{4z}\mathcal{T}$ symmetry remains unaltered (see Note S1 in SI).

\begin{figure}[t]
	\centering
	\subfigure{
		\includegraphics[scale=0.2]{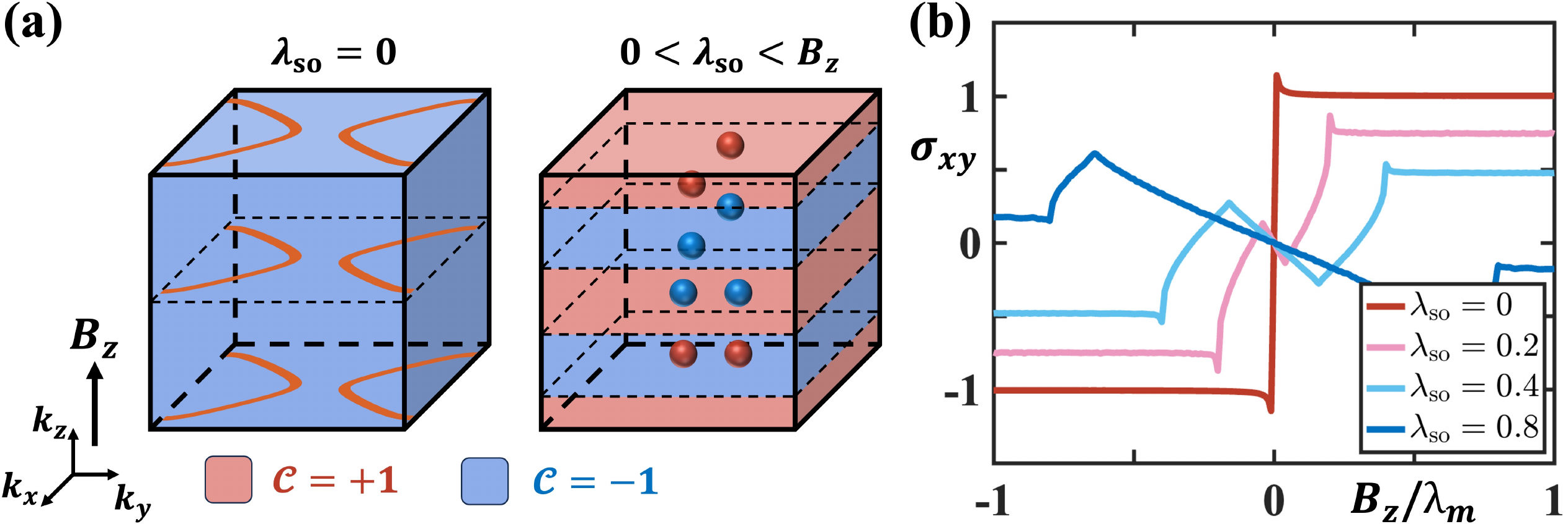}}
	\caption{Anomalous Hall effects activated by a $z$-direction Zeeman field. (a) Two representative band-degeneracy configurations giving rise to Hall plateaus under a Zeeman field applied along the $z$ direction. Left: $\lambda_{\rm so}=0$, where the $z$-direction Zeeman field deforms the CNLs into two nodal rings. Right: $0<\lambda_{\rm so}<B_{z}$, where four pairs of Weyl nodes (spheres in red and blue) remain whose $k_{z}$ components are fixed and independent of $B_{z}$. In both panels, the light red (light blue) regions mark the range of $k_z$ for which each $k_z$ slice is a Chern insulator with $\mathcal{C}=1$ ($\mathcal{C}=-1$).
		(b) Evolution of the Hall conductivity tensor $\sigma_{xy}$ (in units of $e^{2}/h$) as a function of $B_{z}$. 
		Parameters are $t=t_{z}=0$, $\mu=0$, and $\lambda_{\rm M}=\eta=1$.}
	\label{Fig4}
\end{figure}

{\it Anomalous Hall effect.---}As aforementioned, although the Berry curvature is divergently enhanced near band degeneracies, 
the anomalous Hall effect arising from the Berry curvature is entirely canceled due to symmetry constraints. 
This scenario changes once a Zeeman field is applied along the $z$-direction, represented by $B_{z}\sigma_{z}$. 
The Zeeman field breaks the symmetries $\mathcal{M}_{xy}$, $\mathcal{M}_{\bar{x}y}$ and $C_{4z}\mathcal{T}$, thereby allowing a large Hall conductivity $\sigma_{xy}$  regardless of whether SOC
is present~(see Note S3 in SI). In the absence of SOC ($\lambda_{\rm so}=0$), the Zeeman field deforms the CNLs into two nodal rings located at the $k_{z}=0$ and 
$\pi$ planes, protected by the remaining mirror symmetry $\mathcal{M}_{z}$, as shown in 
the left panel of Fig.~\ref{Fig4}(a). All other $k_{z}$ slices are gapped, 
and their Chern numbers are $\mathcal{C}(k_{z})=+1$ ($-1$) for $B_{z}<0$ ($B_{z}>0$), leading to a three-dimensional quantum anomalous Hall effect when the Fermi surface coincides with the two nodal rings~\cite{Lim2017NLSM}.

When $\lambda_{\rm so}$ is finite and $\lambda_{\rm so}<\lambda_{\rm M}$, we find 
that Hall plateaus can also emerge provided the conditions $B_{z}>\lambda_{\rm so}$ and 
$\eta>\eta_{c}+|B_{z}|/2\lambda_{\rm M}$ are met. 
Under these parameter conditions, those Weyl nodes whose $k_{z}$-components of the positions vary with $B_{z}$ annihilate in pairs, leaving behind eight Weyl nodes 
whose $k_{z}$-components are fixed and independent of $B_{z}$. These residual nodes form four pairs, located at momentum planes 
$k_{z}=\{\pm k_{c}$, $\pm (\pi-k_{c})\}$ with $k_{c}=\arcsin(\lambda_{\rm so}/\lambda_{\rm M})$,
as illustrated in the right panel of Fig.~\ref{Fig4}(a). They delineate gapped
$k_{z}$ planes with $\mathcal{C}=1$ from those with $\mathcal{C}=-1$, producing a Hall plateau with quantized value
$\sigma_{xy}=(1-\frac{4}{\pi}\arcsin\frac{\lambda_{\rm so}}{\lambda_{\rm M}})e^{2}/h$~\cite{Yang2011hall,Burkov2011}, as depicted in Fig.~\ref{Fig4}(b). 
We emphasize that this mechanism is different from the conventional realization of three-dimensional quantum Hall effect based on Landau levels, which typically requires a strong magnetic field~\cite{Halperin1987,Bernevig2007QHE,Tang2019QHE}.

\begin{figure}[t]
	\centering
	\subfigure{
		\includegraphics[scale=0.28]{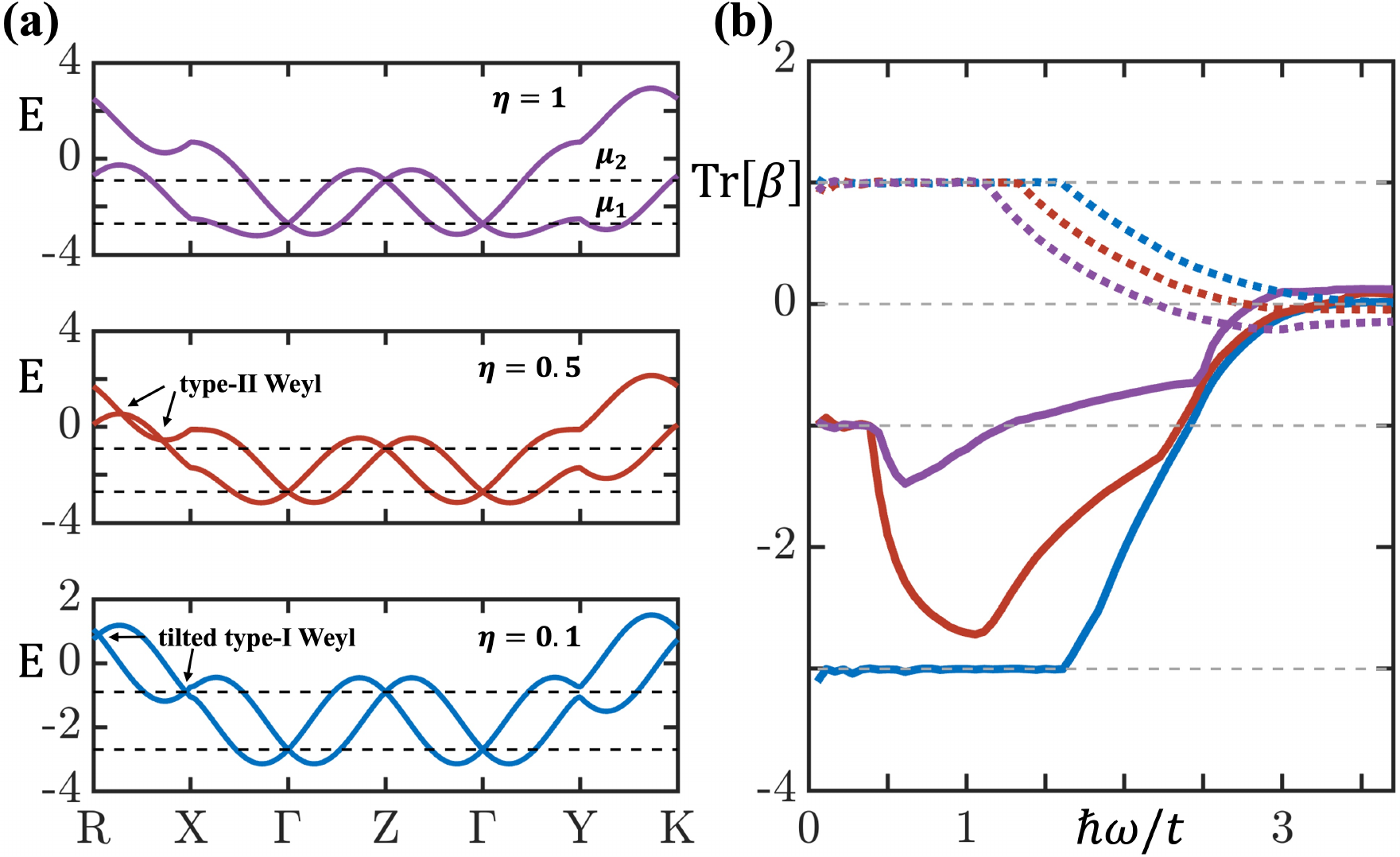}}
	\caption{Quantized circular photogalvanic effect. (a) Band structures along high-symmetry lines of the Brillouin zone for different values of anisotropy parameter $\eta$ in the magnetic exchange field. The two choices of Fermi energy are $\mu_{1}=-2t-t_{z}$ and $\mu_{2}=-2t+t_{z}$. 
		(b) Trace of the CPGE tensor  $\mathrm{Tr}[\beta(\omega)]$ (in units of $i\beta_{0}$).  
		Dashed (solid) curves correspond to the Fermi energy set to $\mu_{1}$ ($\mu_{2}$), with colors matching the corresponding band structures (and $\eta$) in [(a)].
		Common parameters are $t=t_{z}=0.9$, $\lambda_{\rm so}=1$, and $\lambda_{\rm M}=0.8$.}
	\label{Fig5}
\end{figure}

{\it Quantized circular photogalvanic effect.---}When Weyl nodes of opposite topological charges 
are energetically separated and form type-I Weyl cones, they can induce a quantized circular photogalvanic effect (CPGE) within a suitable frequency window~\cite{Juan2017,Chang2017CPGE,Flicker2018CPGE}. Specifically, 
when the momentum-space surface $\bS$, formed by the momenta involved in the optical transition, 
encloses Weyl nodes with net topological charge $\sum_{\boldsymbol q}\mathcal{C}_{\boldsymbol q}$, the trace of the CPGE tensor becomes quantized~\cite{Juan2017}, i.e.,
\begin{equation}
	{\rm Tr}\left[\beta(\omega)\right]=i\frac{e^3}{2h^2}\oint \bOmega \cdot d\bS=i\beta_{0}\sum_{\boldsymbol q}\mathcal{C}_{\boldsymbol q},
\end{equation}
where $\beta_{0}=\pi e^{3}/h^2$. In our system, the presence of SOC breaks inversion and all mirror symmetries, resulting in MKWNs of opposite topological charge being separated in energy, as illustrated in Fig.~\ref{Fig5}(a). 
Additionally, the MKWNs are untilted, suggesting that this system is ideal for the observation of quantized CPGE. 

Remarkably, our system can support CPGEs with enhanced quantization values 
even though $\mathcal{C}_{\boldsymbol q}=\pm1$ for individual Weyl nodes. In Kramers Weyl semimetals, ${\rm Tr}\left[\beta(\omega)\right]$ typically quantizes to $\pm i\beta_{0}$, since time-reversal symmetry does not relate different Kramers Weyl nodes~\cite{Chang2018,Hasan2021}.
In contrast, here the $C_{4z}\mathcal{T}$ symmetry ensures that the Weyl nodes near 
high-symmetry points---such as $\rm X$ and $\rm Y$, or $\rm R$ and $\rm K$---are energetically aligned, enabling higher quantization values such as $\pm2 i\beta_{0}$. In the isotropic limit 
where $t=t_{z}$, even larger quantization like $\pm3 i\beta_{0}$ can be achieved, as illustrated in 
Fig.~\ref{Fig5}(b).

{\it Discussions and conclusions.---}We unveil that CNLs and MKWNs 
are two distinctive forms of band degeneracies that can naturally emerge in spin-split antiferromagnets. 
These band degeneracies are characterized not only by divergent Berry curvature but also by topological boundary states 
exhibiting unconventional patterns of spin textures. Importantly, they give rise to striking physical consequences, including strong---potentially quantized---anomalous Hall effects driven 
by weak Zeeman fields, as well as CPGEs with enhanced quantization.

In the absence of SOC, our model Hamiltonian exhibits band degeneracies consistent with those reported in MnTe$_{2}$\cite{Zhu2024} upon performing a coordinate transformation, suggesting MnTe$_{2}$ may serve as a potential candidate material for observing CNL-related physics (see Note S5 in SI). In addition, chiral antiferromagnets, such as Mn$_{3}$IrGe and YMnO$_{3}$~\cite{Gao2023Heesch}, may provide promising platforms for the realization of MKWNs. These distinctive forms of spin degeneracies can be directly probed by spin-resolved ARPES~\cite{Zhu2024}, and the associated large anomalous Hall response is expected to manifest as an enhanced Kerr rotation angle in magneto-optical Kerr effect (MOKE) measurements~\cite{Zhu2025Alter}. Finally, we note that because the formation of CNLs only requires three mutually orthogonal mirror planes, exact orthogonal CNLs naturally arise in cubic, tetragonal, and orthorhombic crystals. However, similar nodal degeneracies may also appear along high-symmetry lines in hexagonal and trigonal crystals, albeit in generally non-orthogonal configurations. Therefore, the mechanisms we identify are broadly applicable across five of the seven crystal systems. Overall, our findings 
highlight spin-split antiferromagnets as a rich platform for exploring unconventional topological phases 
and the associated phenomena.

{\it Acknowledgements.---}
Z.-Y.Z. and Z.Y. would like to thank Chang Liu, Song-Bo Zhang and Jin-Xin Hu for helpful discussion. We also want to thank the anonymous Referees for their insightful comments that help us improve the paper.
This work was funded by the National Natural Science Foundation of China via grant No. 12174455, No. 12474264, Guangdong Basic and Applied Basic Research Foundation via grant No. 2023B1515040023, Guangdong Provincial Quantum Science Strategic Initiative via grant No. GDZX2404007, and National Key R\&D Program of China via grant No. 2022YFA1404103.

\bibliography{dirac.bib}

\begin{widetext}
\clearpage
\begin{center}
\textbf{\large Supplemental Material for ``Cartesian Nodal Lines and Magnetic Kramers Weyl Nodes in Spin-Split Antiferromagnets''}\\
\vspace{4mm}
{Zheng-Yang Zhuang,$^{1}$ Di Zhu,$^{1}$ Zhigang Wu,$^{2}$ Zhongbo Yan$^{1,*}$}\\
\vspace{2mm}
{\em $^{1}$Guangdong Provincial Key Laboratory of Magnetoelectric Physics and Devices, \\State Key Laboratory of Optoelectronic Materials and Technologies,\\School of Physics, Sun Yat-sen University, Guangzhou 510275, China}\\
{\em $^{2}$Quantum Science Center of Guangdong-Hong Kong-Macao Greater Bay Area (Guangdong), 
	Shenzhen 508045, China}\\
\end{center}

\setcounter{equation}{0}
\setcounter{figure}{0}
\setcounter{table}{0}
\makeatletter
\renewcommand{\theequation}{S\arabic{equation}}
\renewcommand{\thefigure}{S\arabic{figure}}
\renewcommand{\bibnumfmt}[1]{[S#1]}

The supplemental material contains five sections: (I) Band degeneracies in the spin-split band structure; (II) Realization of 
the tight-binding Hamiltonian; (III) Anomalous Hall effects induced by a Zeeman field; (IV) Zeeman field effects on the local magnetic moments; (V) Connection to real-world materials.

\section{I. Band degeneracies in the spin-split band structure}
We start from the effective tight-binding Hamiltonian shown in Eq.~(1) of the main text,
\begin{eqnarray}
	\mathcal{H}(\bk)&=&\varepsilon_{0}(\bk)\sigma_{0}+\lambda_{\rm so}\bl(\bk)\cdot\bsigma+\lambda_{\rm M}\boldm(\bk)\cdot\bsigma\nonumber\\
	&=&\left[-t(\cos k_{x}+\cos k_{y})-t_{z}\cos k_{z}\right]\sigma_{0}+\sin k_{x}(\lambda_{\rm so}-\lambda_{\rm M}\sin k_{z})\sigma_{x}\nonumber\\
	&&+\sin k_{y}(\lambda_{\rm so}+\lambda_{\rm M}\sin k_{z})\sigma_{y}+\left[\lambda_{\rm so}\sin k_{z}+\eta\lambda_{\rm M}(\cos k_{x}-\cos k_{y})\right]\sigma_{z}.
	\label{eq: explicitH}
\end{eqnarray}
The energy spectra are given by
\begin{eqnarray}
	E_{\pm}(\bk)&=&\pm\sqrt{\sin^{2}k_{x}(\lambda_{\rm so}-\lambda_{\rm M}\sin k_{z})^{2}+\sin^{2}k_{y}(\lambda_{\rm so}+\lambda_{\rm M}\sin k_{z})^{2}+\left[\lambda_{\rm so}\sin k_{z}+\eta\lambda_{\rm M}(\cos k_{x}-\cos k_{y})\right]^{2}}\nonumber\\
	&&+\left[-t(\cos k_{x}+\cos k_{y})-t_{z}\cos k_{z}\right].
\end{eqnarray}
Band degeneracies occur at those $\bk$ that simultaneously satisfy the following three conditions: (I) 
$\sin k_{x}(\lambda_{\rm so}-\lambda_{\rm M}\sin k_{z})=0$; (II) $\sin k_{y}(\lambda_{\rm so}+\lambda_{\rm M}\sin k_{z})=0$;
(III) $\lambda_{\rm so}\sin k_{z}+\eta\lambda_{\rm M}(\cos k_{x}-\cos k_{y})=0$. 

In three dimensions, the antisymmetric Berry curvature tensor has three independent components. For this two-band Hamiltonian, 
these components are determined by 
\begin{eqnarray}
	\Omega^{(c)}_{l}(\bk)=-\Omega^{(v)}_{l}(\bk)=\epsilon_{ijl}\frac{\bd(\bk)\cdot(\partial_{i}\bd(\bk)\times\partial_{j}\bd(\bk))}{4|\bd(\bk)|^{3}},
\end{eqnarray}
where the superscript $c/v$ represents conduction/valence band. The term $\epsilon_{ijl}$ represents the antisymmetric Levi-Civita symbol,
where $i$, $j$, and $l$ are indices belonging to the set $\{x,y,z\}$. Additionally, summation over repeated indices is implied. 
The {\zy vector} $\bd(\bk)=\lambda_{\rm so}\bl(\bk)+\lambda_{\rm M}\boldm(\bk)$
with $\bl(\bk)=(\sin k_{x},\sin k_{y},\sin k_{z})$ and $\boldm(\bk)=(-\sin k_{x}\sin k_{z},\sin k_{y}\sin k_{z},\eta(\cos k_{x}-\cos k_{y}))$.

\subsection{A. Cartesian nodal lines}
We first consider the case where spin-orbit coupling (SOC) is absent, i.e., $\lambda_{\rm so}=0$. 
In this case, the Hamiltonian reduces to
\begin{eqnarray}
	\mathcal{H}_{\rm CNL}(\bk)&=&\left[-t(\cos k_{x}+\cos k_{y})-t_{z}\cos k_{z}\right]\sigma_{0}-\lambda_{\rm M}\sin k_{x}\sin k_{z}\sigma_{x}
	+\lambda_{\rm M}\sin k_{y}\sin k_{z}\sigma_{y}\nonumber\\
	&&+\eta\lambda_{\rm M}(\cos k_{x}-\cos k_{y})\sigma_{z},
	\label{eq: CNLSMsH}
\end{eqnarray}
and the energy spectra become
\begin{eqnarray}
	E_{\pm}(\bk)&=&\left[-t(\cos k_{x}+\cos k_{y})-t_{z}\cos k_{z}\right]\pm\lambda_{\rm M}\sqrt{\sin^{2}k_{z}(\sin^{2}k_{x}+\sin^{2}k_{y})+\eta^{2}(\cos k_{x}-\cos k_{y})^{2}}.
\end{eqnarray}
It is straightforward to find that band degeneracies appear at these positions: (1) along the momentum lines satisfying $|k_{x}|=|k_{y}|$ within the $k_{z}=0/\pi$ planes, and (2) along the high-symmetry $k_{z}$ lines passing through $(k_{x},k_{y})=(0/\pi, 0/\pi)$.
These nodal lines intersect orthogonally at the four $C_{4z}\mathcal{T}$-invariant momenta within the Brillouin zone, whose explicit positions are 
at $(0,0,0)$, $(0,0,\pi)$, $(\pi,\pi,0)$,
and $(\pi,\pi,\pi)$. At each intersection, 
the nodal-line structure is analogous to the Cartesian coordinate system, thereby we refer to these nodal lines as 
Cartesian nodal lines (CNLs). The three components of the Berry curvature for this case are given by
\begin{eqnarray}
	\Omega_{x}^{(c)}(\bk)&=&-\Omega_{x}^{(v)}(\bk)=\frac{\eta\sin k_{x}\sin k_{z}\cos k_{y}\cos k_{z}(\cos k_{x}-\cos k_{y})}{2[\sin^{2}k_{z}(\sin^{2}k_{x}+\sin^{2}k_{y})+\eta^{2}(\cos k_{x}-\cos k_{y})^{2}]^{3/2}},\nonumber\\
	\Omega_{y}^{(c)}(\bk)&=&-\Omega_{y}^{(v)}(\bk)=\frac{\eta\sin k_{y}\sin k_{z}\cos k_{x}\cos k_{z}(\cos k_{x}-\cos k_{y})}{2[\sin^{2}k_{z}(\sin^{2}k_{x}+\sin^{2}k_{y})+\eta^{2}(\cos k_{x}-\cos k_{y})^{2}]^{3/2}},\nonumber\\
	\Omega_{z}^{(c)}(\bk)&=&-\Omega_{z}^{(v)}(\bk)=\frac{\eta\sin^{2} k_{z}(\cos k_{x}-\cos k_{y})}{2[\sin^{2}k_{z}(\sin^{2}k_{x}+\sin^{2}k_{y})+\eta^{2}(\cos k_{x}-\cos k_{y})^{2}]^{3/2}}.\label{BC}
\end{eqnarray}
In three dimensions, the antisymmetric Hall conductivity tensor also has three independent components: $\sigma_{xy}$, 
$\sigma_{yz}$ and $\sigma_{zx}$. When only considering 
the contribution from the Berry curvature, their relation with the Berry curvature is given by~\cite{Xiao2010review} 
\begin{eqnarray}
	\sigma_{ij}=\frac{e^{2}}{\hbar}\sum_{n}\int\frac{d^{3}k}{(2\pi)^{3}}\epsilon_{ijl}\Omega_{l}^{(n)}(\bk)f(E_{n}(\bk)).
\end{eqnarray} 
Here, $n$ denotes the band index, and $f(E_{n})=\frac{1}{1+e^{(E_{n}-\mu)/k_{B}T}}$ is the Fermi-Dirac distribution function, where $\mu$ is 
the chemical potential, $k_{B}$ is the Boltzmann constant, and $T$ denotes the temperature.  According to the expressions given in 
Eq.~(\ref{BC}), it is evident that all three components of the Hall conductivity tensor vanish identically. 

The existence of topological boundary states in a nodal-line semimetal is generally characterized 
by a quantized $\pi$ Berry phase or winding number defined on lines traversing the Brillouin zone, provided 
that the considered line has inversion symmetry or chiral symmetry. These two symmetries also provide an intuitive 
understanding of the topological boundary states in this system. For instance, if we consider open boundary conditions 
in the principal $x$ direction, whether there exist topological boundary states can be determined by analyzing 
the reduced one-dimensional Hamiltonian, 
\begin{eqnarray}
	\mathcal{H}_{\rm CNL}(k_{x})=-\lambda_{1}(k_{z})\sin k_{x}\sigma_{x}+\lambda_{2}(k_{y},k_{z})\sigma_{y}+[\eta\cos k_{x}-\eta(k_{y})]\sigma_{z}.
\end{eqnarray}
Here, $\lambda_{1}(k_{z})=\lambda_{\rm M}\sin k_{z}$, $\lambda_{2}(k_{y},k_{z})=\lambda_{\rm M}\sin k_{y}\sin k_{z}$, 
and $\eta(k_{y})=\eta\cos k_{y}$; We have omitted the term $\varepsilon_{0}(\bk)\sigma_{0}$ 
since it only affects the dispersion but has no impact on the existence of topological boundary states;
Furthermore, both $k_{y}$ and $k_{z}$ have been treated as parameters since 
they are good quantum numbers when considering topological boundary states on the $x$-normal surfaces. 
It is evident that the one-dimensional Hamiltonian $\mathcal{H}(k_{x})$ has neither inversion 
symmetry nor chiral symmetry. This simple fact indicates that the Hamiltonian does not have topological 
boundary states on the $x$-normal surfaces. The absence of topological boundary states on the $x$-normal surfaces 
can also be understood by noting that, 
when the CNLs are projected along the $x$ direction, there always exist two nodal lines whose projections overlap (see 
the nodal-line structure shown in Fig.~1 of the main text). 
However, away from the principal axis direction, the projection of certain nodal lines no longer overlaps 
with those of others, leading to the emergence of topological surface states. Below we consider open boundary conditions 
along the $(\bar{1}10)$ direction as an illustrative example. In this case, {\zb the projections of these nodal lines---$(0,0,k_{z})$, $(\pi,\pi,k_{z})$, $(k,k,\pi)$ and $(k,k,0)$---do not overlap with each other}. To determine the topological surface states of this case, 
we perform a coordinate transformation to simplify the analysis. Specifically, we rotate the $k_{x}$-$k_{y}$ plane about the $(0,0,k_{z})$ axis by 
$\pi/4$.  Introduce $k_{1}=\frac{k_{x}-k_{y}}{\sqrt{2}}$, $k_{2}=\frac{k_{x}+k_{y}}{\sqrt{2}}$. The 
Hamiltonian in the new coordinate system reads 
\begin{eqnarray}
	\mathcal{H}_{\rm CNL}(\bk)&=&-\lambda_{\rm M}\sin k_{z}\sin\frac{k_{1}+k_{2}}{\sqrt{2}}\sigma_{x}+\lambda_{\rm M}\sin k_{z}\sin\frac{k_{2}-k_{1}}{\sqrt{2}}\sigma_{y}+\lambda_{\rm M}\eta(\cos\frac{k_{1}+k_{2}}{\sqrt{2}}-\cos\frac{k_{2}-k_{1}}{\sqrt{2}})\sigma_{z}\nonumber\\
	&=&-\lambda_{\rm M}
	\left[\sin k_{z}\cos\frac{k_{1}}{\sqrt{2}}\sin\frac{k_{2}}{\sqrt{2}}(\sigma_{x}-\sigma_{y})+\sin k_{z}\sin\frac{k_{1}}{\sqrt{2}}\cos\frac{k_{2}}{\sqrt{2}}(\sigma_{x}+\sigma_{y})+2\eta\sin\frac{k_{1}}{\sqrt{2}}\sin\frac{k_{2}}{\sqrt{2}}\sigma_{z}\right]\nonumber\\
	&=&-\lambda_{\rm M}
	\left[\sqrt{2}\sin k_{z}\cos\frac{k_{1}}{\sqrt{2}}\sin\frac{k_{2}}{\sqrt{2}}\sigma_{-}+\sqrt{2}\sin k_{z}\sin\frac{k_{1}}{\sqrt{2}}\cos\frac{k_{2}}{\sqrt{2}}\sigma_{+}+2\eta\sin\frac{k_{1}}{\sqrt{2}}\sin\frac{k_{2}}{\sqrt{2}}\sigma_{z}\right]\nonumber\\
	&=&-\lambda_{\rm M}\left\{\sqrt{2}\sin k_{z}\cos\frac{k_{1}}{\sqrt{2}}\sin\frac{k_{2}}{\sqrt{2}}\sigma_{-}+\sin\frac{k_{1}}{\sqrt{2}}\Lambda(k_{2},k_{z})\left[\cos\theta(k_{2},k_{z})\sigma_{+}
	+\sin\theta(k_{2},k_{z})\sigma_{z}\right]\right\}\\
	&=&-\sqrt{2}\lambda_{\rm M}\sin k_{z}\cos\frac{k_{1}}{\sqrt{2}}\sin\frac{k_{2}}{\sqrt{2}}\ \sigma_{-}
	-\lambda_{\rm M}\Lambda(k_{2},k_{z})\sin\frac{k_{1}}{\sqrt{2}}\sigma_{g}(k_{2},k_{z}),
	\label{eq: regroupCNL}
\end{eqnarray}
Above, we have defined
$\sigma_{\pm}\equiv\frac{\sigma_{x}\pm\sigma_{y}}{\sqrt{2}}$,
$\Lambda(k_{2},k_{z})=\sqrt{2\sin^{2} k_{z}\cos^{2}\frac{k_{2}}{\sqrt{2}}+4\eta^{2} \sin^{2} \frac{k_{2}}{\sqrt{2}}}$, and
$\theta(k_{2},k_{z})=\arg(\sqrt{2}\sin k_{z}\cos\frac{k_{2}}{\sqrt{2}}+i 2\eta \sin \frac{k_{2}}{\sqrt{2}})$.
We note that this new set of Pauli matrices $\{\sigma_{-},\sigma_{+},\sigma_{z}\}$ satisfies the same algebra as the standard Pauli matrices $\{\sigma_{x},\sigma_{y},\sigma_{z}\}$. Specifically, they obey the anticommutation relation  $\{\sigma_{i},\sigma_{j}\}=2\delta_{ij}$, and  the commutation relation {\zb $[\sigma_{i},\sigma_{j}]=2i\epsilon_{ijl}\sigma_{l}$}, 
where $i$, $j$ and $l\in\{-,+,z\}$. Here, the Levi-Civita symbol $\epsilon_{ijl}$ is defined such that 
$\epsilon_{-+z}=\epsilon_{+z-}=\epsilon_{z-+}=1$ and $\epsilon_{+-z}=\epsilon_{-z+}=\epsilon_{z+-}=-1$. 
Furthermore, in Eq.~(\ref{eq: regroupCNL}), we mathematically group $\cos\theta(k_{2},k_{z})\sigma_{+}+\sin\theta(k_{2},k_{z})\sigma_{z}$ as $\sigma_{g}(k_{2},k_{z})$. It is obvious that $\sigma_{g}(k_{2},k_{z})$ is a unitary matrix that always anticommutes with $\sigma_{-}$. Hence, there must be another unitary matrix $\mathcal{S}(k_{2},k_{z})=-\sin\theta(k_{2},k_{z})\sigma_{+}+\cos\theta(k_{2},k_{z})\sigma_{z}$ that anticommutes with both $\sigma_{g}(k_{2},k_{z})$ and $\sigma_{-}$, completing the Pauli-like algebra. 

When open boundary conditions are applied in the $(\bar{1}10)$ direction, $k_{2}$ and $k_{z}$ can be treated as 
parameters. The existence of topological boundary states can then be determined by analyzing the topological properties of the reduced one-dimensional Hamiltonian along a given $k_{1}$ line. For fixed values of $k_{2}$ and $k_{z}$, the Hamiltonian possesses both inversion symmetry and chiral symmetry. The inversion symmetry operator is given by
$\mathcal{P}=\sigma_{-}$, which satisfies $\mathcal{P}\mathcal{H}_{\rm CNL}(k_{1},k_{2},k_{z})\mathcal{P}^{-1}
=\mathcal{H}_{\rm CNL}(-k_{1},k_{2},k_{z})$. 
The chiral symmetry operator is exactly the
$\mathcal{S}(k_{2},k_{z})=-\sin\theta(k_{2},k_{z})\sigma_{+}+\cos\theta(k_{2},k_{z})\sigma_{z}$, which satisfies 
the anticommutation relation $\{\mathcal{S}(k_{2},k_{z}),\mathcal{H}_{\rm CNL}(k_{1},k_{2},k_{z})\}=0$. It is noteworthy that 
the inversion symmetry operator $\mathcal{P}$ does not depend on $k_{2}$ and $k_{z}$, while the chiral 
symmetry operator $\mathcal{S}(k_{2},k_{z})$ does. If we restore $k_{2}$ and $k_{z}$ as momentum components,
$\mathcal{S}(k_{2},k_{z})$ can no longer be interpreted as a chiral symmetry operator because it depends on momentum. 
This contradicts the requirement for a chiral symmetry operator, which must be a constant unitary and Hermitian operator.
When a unitary and Hermitian operator, which depends on partial momentum components and anticommutes with the Hamiltonian, exists, 
the Hamiltonian is said to possess subchiral symmetry, and the corresponding operator is referred to as a subchiral symmetry operator\cite{Mo2024subchiral}. This concept turns out to be very useful for diagnosing 
the topology of the Hamiltonian on certain dimension-reduced closed manifolds and the properties of the associated 
topological boundary states.

Since the reduced one-dimensional Hamiltonians possess both inversion symmetry and chiral symmetry, 
their topological properties can be characterized by both the quantized Berry phases and the winding number. 
The quantized Berry phase, as a $Z_{2}$ invariant, has a simple relation with the product of the parity eigenvalues 
at momentum $k_{1}=0$ and $k_{1}=\sqrt{2}\pi$. Namely, $e^{i\phi}=\xi(k_{1}=0)\xi(k_{1}=\sqrt{2}\pi)$, where 
$\xi(K)$ denotes the parity eigenvalue of the occupied states at the inversion-invariant momentum $K$.
It is straightforward to obtain that $\xi(k_{1}=0)\xi(k_{1}=\sqrt{2}\pi)=-[\text{sgn}(\sin k_{z}\sin\frac{k_{2}}{\sqrt{2}})]^{2}$
as long as the energy spectrum of the reduced one-dimensional Hamiltonian is fully gapped. Accordingly, it is evident that 
the Berry phase $\phi$ is quantized to $\pi$ as long as $k_{2}\neq\{0,\sqrt{2}\pi\}$ and $k_{z}\neq\{0,\pi\}$. Therefore, 
the condition for the existence of topological surface states in this case is $k_{2}\neq\{0,\sqrt{2}\pi\}$ and $k_{z}\neq\{0,\pi\}$. 
Although inversion symmetry and parity eigenvalues provide a straightforward method to diagnose 
the existence of topological surface states, they cannot offer further information about the 
spin texture of these states. This limitation arises because inversion symmetry is a spatial 
symmetry, and the topological states on a given surface are not eigenstates of the inversion 
symmetry operator. Interestingly, as noted earlier, the reduced one-dimensional Hamiltonians 
also possess chiral symmetry, and their topological properties can be determined by a winding number. 
The winding number is given by\cite{Ryu2010}
\begin{eqnarray}
	W^{(\bar{1}10)}(k_{2},k_{z})=\frac{1}{4\pi i}\int_{-\sqrt{2}\pi}^{\sqrt{2}\pi}dk_{1} {\rm Tr}
	\left[\mathcal{S}(k_{2},k_{z})\mathcal{H}^{-1}_{\rm CNL}(k_{1},k_{2},k_{z})\partial_{k_{1}}\mathcal{H}_{\rm CNL}(k_{1},k_{2},k_{z})\right]. 
\end{eqnarray} 
By straightforward calculations, we find that 
\begin{eqnarray}
	W^{(\bar{1}10)}(k_{2},k_{z})={\rm sgn}(k_{2}k_{z}),
\end{eqnarray}
provided that $k_{2}\neq\{0,\sqrt{2}\pi\}$ and $k_{z}\neq\{0,\pi\}$. It is readily seen that the winding number has two nontrivial values, 
$\pm1$. Compared to the single value $\phi=\pi$, this suggests that the chiral symmetry can provide more information on the topological 
surface states. Indeed, since the chiral symmetry is a nonspatial symmetry, the zero-energy surface states also serve as 
the eigenstates of the chiral symmetry operator. Since the two eigenstates of the operator $\mathcal{S}(k_{2},k_{z})$ are straightforward to obtain, 
the spin texture of the surface states can readily be determined. {\zb Specifically}, since $\mathcal{S}(k_{2},k_{z})=-\sin\theta(k_{2},k_{z})\sigma_{+}+\cos\theta(k_{2},k_{z})\sigma_{z}$, 
it is straightforward to obtain its two eigenstates, which are 
\begin{eqnarray}
	|u_{+}(k_{2},k_{z})\rangle=\left(
	\begin{array}{c}
		\cos\frac{\theta(k_{2},k_{z})}{2} \\
		-e^{i\frac{\pi}{4}}\sin\frac{\theta(k_{2},k_{z})}{2} \\
	\end{array}
	\right), \quad 
	|u_{-}(k_{2},k_{z})\rangle=\left(
	\begin{array}{c}
		\sin\frac{\theta(k_{2},k_{z})}{2} \\
		e^{i\frac{\pi}{4}}\cos\frac{\theta(k_{2},k_{z})}{2} \\
	\end{array}
	\right), 
\end{eqnarray}
where the subscripts $\pm$ indicate that $\mathcal{S}(k_{2},k_{z})|u_{\alpha}(k_{2},k_{z})\rangle=\alpha|u_{\alpha}(k_{2},k_{z})\rangle$.
The spin textures associated with these two eigenstates are 
\begin{eqnarray}
	\braket{\sigma_{z}}_{\alpha}(k_{2},k_{z})&=&\langle u_{\alpha}(k_{2},k_{z})|\sigma_{z}|u_{\alpha}(k_{2},k_{z})\rangle=\alpha\cos\theta(k_{2},k_{z}),\nonumber\\
	\braket{\sigma_{+}}_{\alpha}(k_{2},k_{z})&=&\langle u_{\alpha}(k_{2},k_{z})|\sigma_{+}|u_{\alpha}(k_{2},k_{z})\rangle=-\alpha\sin\theta(k_{2},k_{z}),\nonumber\\
	\braket{\sigma_{-}}_{\alpha}(k_{2},k_{z})&=&\langle u_{\alpha}(k_{2},k_{z})|\sigma_{-}|u_{\alpha}(k_{2},k_{z})\rangle=0,
\end{eqnarray}
or in the original spin basis, 
\begin{eqnarray}
	\braket{\sigma_{z}}_{\alpha}(k_{2},k_{z})&=&\langle u_{\alpha}(k_{2},k_{z})|\sigma_{z}|u_{\alpha}(k_{2},k_{z})\rangle=\alpha\cos\theta(k_{2},k_{z}),\nonumber\\
	\braket{\sigma_{y}}_{\alpha}(k_{2},k_{z})&=&\langle u_{\alpha}(k_{2},k_{z})|\sigma_{y}|u_{\alpha}(k_{2},k_{z})\rangle=-\alpha\frac{\sqrt{2}}{2}\sin\theta(k_{2},k_{z}),\nonumber\\
	\braket{\sigma_{x}}_{\alpha}(k_{2},k_{z})&=&\langle u_{\alpha}(k_{2},k_{z})|\sigma_{x}|u_{\alpha}(k_{2},k_{z})\rangle=-\alpha\frac{\sqrt{2}}{2}\sin\theta(k_{2},k_{z}).
\end{eqnarray}
It is noteworthy that when the zero-energy state on one surface aligns with the positive-eigenvalue 
eigenstate of the chiral symmetry operator, the corresponding zero-energy state on the opposing surface 
aligns with the negative-eigenvalue eigenstate. Furthermore, the eigenvalue of the chiral symmetry 
operator for a zero-energy state on a specific surface is directly related to the winding number. 
Specifically, when the winding number changes sign, the eigenvalue for the zero-energy state on that surface will also change sign.
By analyzing the wave functions of the surface states and considering these facts, we obtain 
the spin textures associated with the topological states on the $(\bar{1}10)$ surfaces, which read
\begin{eqnarray}
	\braket{\sigma_{z}}(k_{2},k_{z})&=&\beta {\rm sgn}(k_{2}k_{z})\cos\theta(k_{2},k_{z}),\nonumber\\
	\braket{\sigma_{+}}(k_{2},k_{z})&=&-\beta {\rm sgn}(k_{2}k_{z})\sin\theta(k_{2},k_{z}),\nonumber\\
	\braket{\sigma_{-}}(k_{2},k_{z})&=&0,
	\label{eq: ST x=1}
\end{eqnarray}
where $\beta=+1$ ($-1$) refers to the left (right) surface. 
These results demonstrate that the spin polarizations of the 
topological surface flat bands are momentum-dependent. This behavior is 
fundamentally different from that of the topological surface flat bands 
in a nodal-line semimetal protected by chiral symmetry, where 
the spin polarizations are fixed and momentum-independent.

\subsection{B. Kramers Weyl nodes}

When $\lambda_{\rm M}=0$ and $\lambda_{\rm so}\neq0$, the Hamiltonian reduces to
\begin{eqnarray}
	\mathcal{H}(\bk)&=&\left[-t(\cos k_{x}+\cos k_{y})-t_{z}\cos k_{z}\right]\sigma_{0}+\lambda_{\rm so}\sin k_{x}\sigma_{x}
	+\lambda_{\rm so}\sin k_{y}\sigma_{y}+\lambda_{\rm so}\sin k_{z}\sigma_{z}.
	\label{eq: KWSM}
\end{eqnarray}
This Hamiltonian describes a Kramers Weyl semimetal \cite{Chang2018,Hasan2021}. Its band structure possesses 
Weyl nodes at every time-reversal invariant momentum (TRIM), a consequence of the Kramers degeneracy enforced by spinful time-reversal symmetry. The time-reversal symmetry operator is given by $\mathcal{T}=i\sigma_{y}\mathcal{K}$, which 
satisfies $\mathcal{T}\mathcal{H}(\bk)\mathcal{T}^{-1}=\mathcal{H}(-\bk)$ and $\mathcal{T}^{2}=-1$, where $\mathcal{K}$ denotes 
the complex conjugation operator. Near these nodes, it is known that the Berry curvature has a monopole-like dependence on the momentum 
measured from the corresponding node\cite{Armitage2018RMP}. That is, 
\begin{eqnarray}
	\bOmega^{(c)}_{(n_{1},n_{2},n_{3})\pi}(\bk)=-\bOmega^{(v)}_{_{(n_{1},n_{2},n_{3})\pi}}(\bk)=(-1)^{n_{1}+n_{2}+n_{3}}\frac{\bk}{2\bk^{3}}.
\end{eqnarray}
Here, $n_{i=1,2,3}\in\{0,1\}$, and the subscript $(n_{1},n_{2},n_{3})\pi$ characterizes the TRIM at which
one Weyl node is located. 
The topological charge of each node is characterized by the Chern number, 
which is defined as an integral of the Berry curvature over a closed surface $\bS$ enclosing the corresponding node, i.e.,
\begin{eqnarray}
	\mathcal{C}_{(n_{1},n_{2},n_{3})\pi}=\frac{1}{2\pi}\oint\bOmega_{(n_{1},n_{2},n_{3})\pi}^{(c)}\cdot d\bS.
\end{eqnarray}
The result is 
\begin{eqnarray}
	\mathcal{C}_{(n_{1},n_{2},n_{3})\pi}=(-1)^{n_{1}+n_{2}+n_{3}}.
	\label{eq: KWSM chirality}
\end{eqnarray}
The result indicates that the Weyl nodes at $(0,0,0)$, $(0,\pi,\pi)$, $(\pi,0,\pi)$ and $(\pi,\pi,0)$ have 
topological charge $\mathcal{C}=1$, and the other Weyl nodes at $(\pi,\pi,\pi),\ (0,0,\pi),\ (0,\pi,0)$ and $(\pi,0,0)$
have topological charge $\mathcal{C}=-1$.

\subsection{C. Magnetic Kramers Weyl nodes}

When both $\lambda_{\rm M}$ and $\lambda_{\rm so}$ are finite, the band degeneracies in the band structure 
remain to be Weyl nodes,  but the distributions of these Weyl nodes becomes a little more complex compared to that of a 
Kramers Weyl semimetal. First, because the $C_{4z}\mathcal{T}$ is preserved in the Hamiltonian, the four $C_{4z}\mathcal{T}$-invariant 
momentums, including $(0,0,0)$, $(0,0,\pi)$, $(\pi,\pi,0)$, and $(\pi,\pi,\pi)$, are the locations of symmetry-enforced Weyl nodes. 
As these $C_{4z}\mathcal{T}$-invariant momentums are also TRIMs, we refer to these position-fixed Weyl nodes as magnetic Kramers 
Weyl nodes (MKWNs),  highlighting their positions at TRIMs and the breaking of time-reversal symmetry.   

To determine the distribution of potential additional Weyl nodes, we divide the analysis into two scenarios: 
one where $\lambda_{\rm M}>\lambda_{\rm so}$, and the other where $\lambda_{\rm M}<\lambda_{\rm so}$. Throughout, we consider 
$\lambda_{\rm M}$, $\lambda_{\rm so}$ and $\eta$ to be positive constants. 
Recall the conditions for the emergence of band degeneracies: (I) 
$\sin k_{x}(\lambda_{\rm so}-\lambda_{\rm M}\sin k_{z})=0$; (II) $\sin k_{y}(\lambda_{\rm so}+\lambda_{\rm M}\sin k_{z})=0$;
(III) $\lambda_{\rm so}\sin k_{z}+\eta\lambda_{\rm M}(\cos k_{x}-\cos k_{y})=0$. We begin by examining the scenario where
$\lambda_{\rm M}<\lambda_{\rm so}$, as it is simpler to analyze. For this case,  condition (I) determines 
$k_{x}=0$ or $\pi$. Similarly, condition (II) determines $k_{y}=0$ or $\pi$. The results obtained under 
these two conditions indicate that the Weyl nodes must be located along the four high-symmetry $k_{z}$ lines at 
$(0,0,k_{z})$, $(0,\pi,k_{z})$, $(\pi,0,k_{z})$ and $(\pi,\pi,k_{z})$. On the two $C_{4z}\mathcal{T}$-invariant lines
at $(0,0,k_{z})$ and $(\pi,\pi,k_{z})$, condition (III) reduces to $\lambda_{\rm so}\sin k_{z}=0$, which 
leads to $k_{z}=0$ and $\pi$, suggesting that only the four MKWNs appear on these two lines.  
On the line at $(0,\pi,k_{z})$,  condition (III) simplifies to: $\lambda_{\rm so}\sin k_{z}+2\eta\lambda_{\rm M}=0$, 
which has solutions only if $\lambda_{\rm so}>2\eta\lambda_{\rm M}$. When this condition is fulfilled, two Weyl nodes 
emerge at $(0,\pi,-k_{0})$ and $(0,\pi,-\pi+k_{0})$, where $k_{0}=\arcsin 2\eta\lambda_{\rm M}/\lambda_{\rm so}$. Similarly, 
on the line at $(\pi,0,k_{z})$, two additional Weyl nodes emerge at $(\pi,0,k_{0})$ and $(\pi,0,\pi-k_{0})$,  
provided that $\lambda_{\rm so}>2\eta\lambda_{\rm M}$. 
A schematic of the these Weyl nodes and their associated Fermi arcs is provided in Fig.~\ref{SFig1}.

\begin{figure}[t]
	\centering
	\subfigure{
		\includegraphics[scale=0.5]{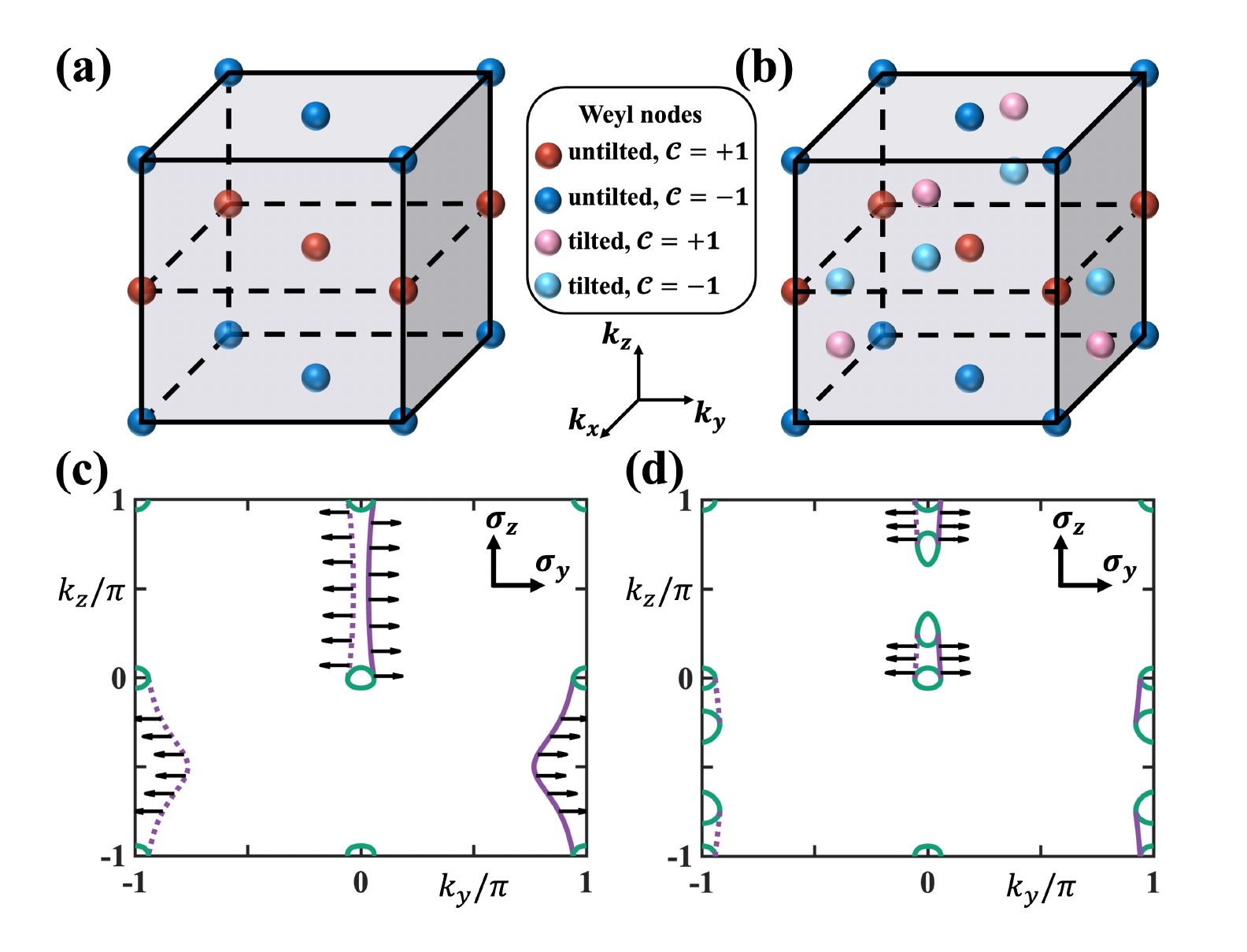}}
		\caption{(a) and $\eta<\eta_{c}$ in (b). Topological charges and cone tilting details of the Weyl nodes are shown in the middle inset. 
			(c-d) Solid (dotted) purple lines refer to Fermi arcs on the left (right) $x$-normal surface, and the black arrows denote their spin polarizations.
			The solid green rings represent the projections of bulk Fermi surface, with the chemical potential fixed at $\mu=0.2$. 
			$\lambda_{\rm M}=0.4$ and $0.8$ in (c) and (d), respectively. Common parameters are $t=t_{z}=0$, $\lambda_{\rm so}=1.1$, and $\eta=1$.}
	\label{SFig1}
\end{figure}

Next, we examine the scenario where $\lambda_{\rm M}>\lambda_{\rm so}$. For this case, 
condition (I) yields the following solutions: 
\begin{eqnarray}
	k_{x}=\{0,\, \pi\}, \quad\text{or}\quad k_{z}=\{\arcsin\frac{\lambda_{\rm so}}{\lambda_{\rm M}},\,\pi-\arcsin\frac{\lambda_{\rm so}}{\lambda_{\rm M}}\}.
\end{eqnarray}
Similarly, condition (II) has the following solutions: 
\begin{eqnarray}
	k_{y}=\{0, \,\pi\}, \quad\text{or}\quad k_{z}=\{-\arcsin\frac{\lambda_{\rm so}}{\lambda_{\rm M}}, \,-\pi+\arcsin\frac{\lambda_{\rm so}}{\lambda_{\rm M}}\}. 
\end{eqnarray}
It is easy to see that besides the Weyl nodes discussed in the first scenario,  additional Weyl nodes may emerge 
along the following lines:
\begin{eqnarray}
	&(k_{x},0,\arcsin\frac{\lambda_{\rm so}}{\lambda_{\rm M}}), \, (k_{x},\pi,\arcsin\frac{\lambda_{\rm so}}{\lambda_{\rm M}}), \,
	(k_{x},0,\pi-\arcsin\frac{\lambda_{\rm so}}{\lambda_{\rm M}}),  \,(k_{x},\pi,\pi-\arcsin\frac{\lambda_{\rm so}}{\lambda_{\rm M}}),\nonumber\\
	&(0,k_{y},-\arcsin\frac{\lambda_{\rm so}}{\lambda_{\rm M}}), \,(\pi,k_{y},-\arcsin\frac{\lambda_{\rm so}}{\lambda_{\rm M}}),\,
	(0,k_{y},-\pi+\arcsin\frac{\lambda_{\rm so}}{\lambda_{\rm M}}), \,(\pi,k_{y},-\pi+\arcsin\frac{\lambda_{\rm so}}{\lambda_{\rm M}}). 
\end{eqnarray}
We now analyze each case individually. 

(1) On the line at $(k_{x},0,\arcsin\frac{\lambda_{\rm so}}{\lambda_{\rm M}})$: Condition (III) reduces to
\begin{eqnarray}
	\frac{\lambda_{\rm so}^{2}}{\lambda_{\rm M}}+\lambda_{\rm M}\eta(\cos k_{x}-1)=0.
\end{eqnarray} 
It has solutions at 
$k_{x}=\pm \arccos\left(1-\frac{\lambda_{\rm so}^{2}}{\lambda_{\rm M}^{2}\eta}\right)$, provided 
that the condition $\eta>\frac{\lambda_{\rm so}^{2}}{\lambda_{\rm M}^{2}}$ is satisfied.

(2) On the line  
$(k_{x},\pi,\arcsin\frac{\lambda_{\rm so}}{\lambda_{\rm M}})$: Condition (III) reduces 
to 
\begin{eqnarray}
	\frac{\lambda_{\rm so}^{2}}{\lambda_{\rm M}}+\lambda_{\rm M}\eta(\cos k_{x}+1)=0,
\end{eqnarray}
which has no solutions. 

(3) On the line $(k_{x},0,\pi-\arcsin\frac{\lambda_{\rm so}}{\lambda_{\rm M}})$: 
Condition (III) reduces to the same equation as in case (1), yielding the same solutions
\begin{eqnarray}
	k_{x}=\pm \arccos\left(1-\frac{\lambda_{\rm so}^{2}}{\lambda_{\rm M}^{2}\eta}\right)
\end{eqnarray}
under the same condition $\eta>\frac{\lambda_{\rm so}^{2}}{\lambda_{\rm M}^{2}}$. 

(4) On the line $(k_{x},\pi,\pi-\arcsin\frac{\lambda_{\rm so}}{\lambda_{\rm M}})$: Condition (III) reduces to the same equation as in case (2),
thereby no solutions exist. 

(5) On the line $(0,k_{y},-\arcsin\frac{\lambda_{\rm so}}{\lambda_{\rm M}})$: The solutions 
are 
\begin{eqnarray}
	k_{y}=\pm \arccos\left(1-\frac{\lambda_{\rm so}^{2}}{\lambda_{\rm M}^{2}\eta}\right).
\end{eqnarray} 

(6) On the line $(\pi,k_{y},-\arcsin\frac{\lambda_{\rm so}}{\lambda_{\rm M}})$: No solutions exist. 

(7) On the line at $(0,k_{y},-\pi+\arcsin\frac{\lambda_{\rm so}}{\lambda_{\rm M}})$: The solutions 
are 
\begin{eqnarray}
	k_{y}=\pm \arccos\left(1-\frac{\lambda_{\rm so}^{2}}{\lambda_{\rm M}^{2}\eta}\right).
\end{eqnarray} 

(8) On the line at $(\pi,k_{y},-\pi+\arcsin\frac{\lambda_{\rm so}}{\lambda_{\rm M}})$: No solutions exist. 

In summary, when $\lambda_{\rm M}>\lambda_{\rm so}$ and $\eta>\frac{\lambda_{\rm so}^{2}}{\lambda_{\rm M}^{2}}$, there are 
additional four pairs of Weyl nodes that emerge at four $k_{z}$ planes. Their positions and topological charges are 
summarized as follows:
\begin{eqnarray}
	(k_{x},k_{y},k_{z})=\left\{\begin{array}{cc}
		(\pm k_{w},0,k_{zw}),\ (0,\pm k_{w},-k_{zw}), & \mathcal{C}=-1,\\
		(\pm k_{w},0,\pi-k_{zw}),\ (0,\pm k_{w},-\pi+k_{zw}), & \mathcal{C}=+1,
	\end{array}\right.
	\label{eq: nodes lmglso}
\end{eqnarray}
where $k_{w}=\arccos\left(1-\frac{\lambda_{\rm so}^{2}}{\lambda_{\rm M}^{2}\eta}\right)$ and $k_{zw}=\arcsin(\lambda_{\rm so}/\lambda_{\rm M})$.

\section{II. Realization of the tight-binding Hamiltonian}

In this section, we present a specific magnetic configuration as an example to 
demonstrate a possible realization of the tight-binding Hamiltonian under consideration. 
The distribution of magnetic and nonmagnetic atoms is presented in Figure 1 {\zy in the main text},
where the blue (white) spheres represent magnetic (nonmagnetic) atoms, and the red arrows 
indicate the orientations of local magnetic moments. We assume that the outer-shell electrons 
of magnetic atoms---responsible for forming local moments---are strongly localized, 
whereas those of nonmagnetic atoms are itinerant and govern the system’s transport properties. 
Because of this separation in localization and delocalization, our model describes the 
behavior of these itinerant electrons moving in the magnetic background formed by these magnetic atoms. 
When the  hopping paths of these itinerant electrons intersect with the localized magnetic moments, they 
experience a spin-dependent potential. This potential, in turn, gives rise to spin-dependent hopping amplitudes 
as electrons with spins parallel or antiparallel to the localized magnetic moment exhibit different tunneling probabilities. For instance, the representative hoppings in the $xy$, $yz$, and $zx$ planes are shown in Figure~1{\zy (c-e) in the main text}. In the $xy$ plane, hopping along the $x$ and $y$ directions becomes spin-polarized as $\pm\sigma_{z}$; while in the $yz$ ($xz$) plane, diagonal hopping paths acquire $\pm\sigma_{y}$ ($\pm\sigma_{x}$) spin dependence.
Key real-space symmetries of this structure includes three mirror symmetries $\mathcal{M}_{z}$, $\mathcal{M}_{xy}$, $\mathcal{M}_{\bar{x}y}$, two {\zb twofold} roto-time-reversal symmetries $C_{2}\mathcal{T}$ about the {\zy principal} axes ($x$, $y$), and a {\zy fourfold roto-time-reversal-symmetry} $C_{4z}\mathcal{T}$. These symmetries dictate the allowed hoppings and band degeneracies of the system. 
Based on the magnetic configuration illustrated in Figure 1 in the main {\zb text}, the tight-binding Hamiltonian describing the itinerant electrons (assuming a single orbital degree of freedom for these electrons) is given by

\begin{eqnarray}
	H&=&-\sum_{\braket{i,j},\sigma}t_{ij}c_{i,\sigma}^{\dagger}c_{j,\sigma}
	+i\lambda_{\rm so}\sum_{\braket{i,j},\sigma,\sigma^{'}}\bd_{ij}\cdot c_{i,\sigma}^{\dagger}\bsigma_{\sigma\sigma^{'}}c_{j,\sigma^{'}}\nonumber\\
	&&+\lambda_{\rm M}\eta\sum_{\braket{i,j},\sigma\sigma^{'}}\bS_{ij}\cdot c_{i,\sigma}^{\dagger}\bsigma_{\sigma\sigma^{'}}c_{j,\sigma^{'}}
	+\lambda_{\rm M}\sum_{\braket{\braket{i,j}},\sigma\sigma^{'}}\bS_{ij}\cdot c_{i,\sigma}^{\dagger}\bsigma_{\sigma\sigma^{'}}c_{j,\sigma^{'}}+h.c.\nonumber\\
	&=&\sum_{\bk}\Psi_{\bk}^{\dagger}\mathcal{H}(\bk)\Psi_{\bk}.
	\label{eq: real space H}
\end{eqnarray}
Here, $c_{i,\sigma}(c_{i,\sigma}^{\dagger})$ represents the annihilation (creation) operator for an electron with spin $\sigma$ at site $i$.
The notation $\braket{i,j}$ indicates nearest-neighbor hopping between sites $i$ and $j$, $\braket{\braket{i,j}}$ indicates next-nearest-neighbor 
hopping, and the unit vector $\bd_{ij}$ points along the bond direction from site $j$ to site $i$. The parameter $t_{ij}$ refers to the hopping amplitude between two nearest-neighbor sites, and $\lambda_{\rm M}$ characterizes the difference in hopping amplitude for opposite spins that is induced by the background magnetic moments. {\zy $\bS_{ij}$ denotes the local magnetic moment along the hopping path between site $i$ and $j$.} The spin-orbit coupled terms of the amplitude $\lambda_{\rm so}$ {\zy arise when there are} perturbations that break the inversion and three mirror symmetries. By performing a Fourier transformation 
to the momentum space and choosing the basis as $\Psi_{\bk}^{\dagger}=(c_{\uparrow,\bk}^{\dagger}, c_{\downarrow,\bk}^{\dagger})$, we obtain the momentum-space Hamiltonian, which reads
\begin{eqnarray}
	\mathcal{H}(\bk)&=&-\left[t(\cos k_{x}+\cos k_{y})+t_{z}\cos k_{z}\right]+\lambda_{\rm so}(\sin k_{x}\sigma_{x}+\sin k_{y}\sigma_{y}+\sin k_{z}\sigma_{z})\nonumber\\
	&&+\lambda_{\rm M}\eta(\cos k_{x}-\cos k_{y})\sigma_{z}+\lambda_{\rm M}\sin k_{z}(-\sin k_{x}\sigma_{x}+\sin k_{y}\sigma_{y}),
	\label{eq: effH}
\end{eqnarray}
where the Pauli matrices $(\sigma_{x},\sigma_{y},\sigma_{z})$ acts on the spin degrees of freedom. Importantly, our model Eq.~(\ref{eq: real space H}) includes all symmetry-allowed nontrivial terms from nearest- and next-nearest-neighbor hoppings. Longer-range spin-independent contributions---such as $\sin k_{z}(\sin k_{x}+\sin k_{y})\sigma_{0}$ terms---are omitted because they do not affect the system's topology. Those longer-range spin-dependent terms---such as $(\cos 2k_{x}-\cos2k_{y})\sigma_{z}$---are also neglected since their {\zy amplitudes} are expected to be negligible within the tight-binding {\zy framework} and do not qualitatively alter the results.

\begin{table}[t]
	\centering
	\begin{tabular}{|c|c|c|}
		\hline
		Weyl node position & Charge &  Requirements for {\zy existence}\\
		\hline
		\multirow{2}{*}{$\left(0,0,-\arcsin \frac{B_{z}}{\lambda_{\rm so}}\right)$}
		&\multirow{2}{*}{$+1$}
		&\multirow{8}{*}{$|B_{z}|<\lambda_{\rm so}$}\\
		&&\\
		\multirow{2}{*}{$\left(0,0,-\pi+\arcsin\frac{B_{z}}{\lambda_{\rm so}}\right)$}
		&\multirow{2}{*}{$-1$}&\\
		&&\\
		\cline{1-2}
		\multirow{2}{*}{$\left(\pi,\pi,-\arcsin \frac{B_{z}}{\lambda_{\rm so}}\right)$}
		&\multirow{2}{*}{$+1$}&\\
		&& \\
		\multirow{2}{*}{$\left(\pi,\pi,-\pi+\arcsin\frac{B_{z}}{\lambda_{\rm so}}\right)$}
		&\multirow{2}{*}{$-1$}&\\
		&&\\
		\hline
		\multirow{2}{*}{$\left(0,\pi,-\arcsin\frac{B_{z}+2\lambda_{\rm M}\eta}{\lambda_{\rm so}}\right)$}
		&\multirow{2}{*}{$-1$}
		&\multirow{4}{*}{$\eta<\eta_{c}-\frac{B_{z}}{2\lambda_{\rm M}}$}\\
		&&\\
		\multirow{2}{*}{$\left(0,\pi,-\pi+\arcsin\frac{B_{z}+2\lambda_{\rm M}\eta}{\lambda_{\rm so}}\right)$}
		&\multirow{2}{*}{$+1$}&\\
		&&\\
		\hline
		\multirow{2}{*}{$\left(\pi,0,\arcsin\frac{2\lambda_{\rm M}\eta-B_{z}}{\lambda_{\rm so}}\right)$}
		&\multirow{2}{*}{$-1$}
		&\multirow{4}{*}{$\eta<\eta_{c}+\frac{B_{z}}{2\lambda_{\rm M}}$}\\
		&&\\
		\multirow{2}{*}{$\left(\pi,0,\pi-\arcsin\frac{2\lambda_{\rm M}\eta-B_{z}}{\lambda_{\rm so}}\right)$}
		&\multirow{2}{*}{$+1$}&\\
		&&\\
		\hline
		\multirow{2}{*}{$\left(\pm\arccos(1-\frac{\lambda_{\rm so}^{2}}{\lambda_{\rm M}^{2}\eta}-\frac{B_{z}}{\lambda_{\rm M}\eta}),0,\arcsin\frac{\lambda_{\rm so}}{\lambda_{\rm M}}\right)$}
		&\multirow{2}{*}{$-1$}
		&\multirow{4}{*}{\makecell{$\lambda_{\rm so}<\lambda_{\rm M}$,\\$-\frac{\lambda_{\rm so}^{2}}{\lambda_{\rm M}}<B_{z}<\lambda_{\rm M}(2\eta-\frac{\lambda_{\rm so}^{2}}{\lambda_{\rm M}^{2}})$}}\\
		&&\\
		\multirow{2}{*}{$\left(\pm\arccos(1-\frac{\lambda_{\rm so}^{2}}{\lambda_{\rm M}^{2}\eta}-\frac{B_{z}}{\lambda_{\rm M}\eta}),0,\pi-\arcsin\frac{\lambda_{\rm so}}{\lambda_{\rm M}}\right)$}
		&\multirow{2}{*}{$+1$}&\\
		&&\\
		\hline
		\multirow{2}{*}{$\left(\pm\arccos(1+\frac{\lambda_{\rm so}^{2}}{\lambda_{\rm M}^{2}\eta}+\frac{B_{z}}{\lambda_{\rm M}\eta}),\pi,\arcsin\frac{\lambda_{\rm so}}{\lambda_{\rm M}}\right)$}
		&\multirow{2}{*}{$-1$}
		&\multirow{4}{*}{\makecell{$\lambda_{\rm so}<\lambda_{\rm M}$,\\$-\lambda_{\rm M}(2\eta+\frac{\lambda_{\rm so}^{2}}{\lambda_{\rm M}^{2}})<B_{z}<-\frac{\lambda_{\rm so}^{2}}{\lambda_{\rm M}}$}}\\
		&&\\
		\multirow{2}{*}{$\left(\pm\arccos(1+\frac{\lambda_{\rm so}^{2}}{\lambda_{\rm M}^{2}\eta}+\frac{B_{z}}{\lambda_{\rm M}\eta}),\pi,\pi-\arcsin\frac{\lambda_{\rm so}}{\lambda_{\rm M}}\right)$}
		&\multirow{2}{*}{$+1$}&\\
		&&\\
		\hline
		\multirow{2}{*}{$\left(\pi,\pm\arccos(1+\frac{\lambda_{\rm so}^{2}}{\lambda_{\rm M}^{2}\eta}-\frac{B_{z}}{\lambda_{\rm M}\eta}),-\arcsin\frac{\lambda_{\rm so}}{\lambda_{\rm M}}\right)$}
		&\multirow{2}{*}{$-1$}
		&\multirow{4}{*}{\makecell{$\lambda_{\rm so}<\lambda_{\rm M}$,\\$\frac{\lambda_{\rm so}^{2}}{\lambda_{\rm M}}<B_{z}<\lambda_{\rm M}(2\eta+\frac{\lambda_{\rm so}^{2}}{\lambda_{\rm M}^{2}})$}}\\
		&&\\
		\multirow{2}{*}{$\left(\pi,\pm\arccos(1+\frac{\lambda_{\rm so}^{2}}{\lambda_{\rm M}^{2}\eta}-\frac{B_{z}}{\lambda_{\rm M}\eta}),-\pi+\arcsin\frac{\lambda_{\rm so}}{\lambda_{\rm M}}\right)$}
		&\multirow{2}{*}{$+1$}&\\
		&&\\
		\hline
		\multirow{2}{*}{$\left(0,\pm\arccos(1-\frac{\lambda_{\rm so}^{2}}{\lambda_{\rm M}^{2}\eta}+\frac{B_{z}}{\lambda_{\rm M}\eta}),-\arcsin\frac{\lambda_{\rm so}}{\lambda_{\rm M}}\right)$}
		&\multirow{2}{*}{$-1$}
		&\multirow{4}{*}{\makecell{$\lambda_{\rm so}<\lambda_{\rm M}$,\\$-\lambda_{\rm M}(2\eta-\frac{\lambda_{\rm so}^{2}}{\lambda_{\rm M}^{2}})<B_{z}<\frac{\lambda_{\rm so}^{2}}{\lambda_{\rm M}}$}}\\
		&&\\
		\multirow{2}{*}{$\left(0,\pm\arccos(1-\frac{\lambda_{\rm so}^{2}}{\lambda_{\rm M}^{2}\eta}+\frac{B_{z}}{\lambda_{\rm M}\eta}),-\pi+\arcsin\frac{\lambda_{\rm so}}{\lambda_{\rm M}}\right)$}
		&\multirow{2}{*}{$+1$}&\\
		&&\\
		\hline
	\end{tabular}
	\caption{{\bf Positions, charges, and existence conditions for Weyl nodes under a  Zeeman field in the $z$ direction.} 
		$\eta_{c}=\lambda_{\rm so}/2\lambda_{\rm M}$.}
	\label{table: Weyl}
\end{table}

\section{III. Anomalous Hall effect induced by a Zeeman field}

Before proceeding, we demonstrate that the conservation of  $C_{4z}\mathcal{T}$ symmetry
forces all components of the Hall conductivity tensor to vanish identically. To illustrate this, 
we first examine the component $\sigma_{xy}$. From Ohm's law, the relationship between the current component and the electric field component is given by
\begin{eqnarray}
	j_{x}=\sigma_{xy}E_{y},\quad j_{y}=\sigma_{yx}E_{x},\label{Ohm1}
\end{eqnarray}
where $j_{x}$ and $j_{y}$ represent the current components along the $x$ and $y$ directions, respectively,
while $E_{x}$ and $E_{y}$ represent  the electric field components
in the $x$ and $y$ directions. Under the $C_{4z}$ operation, their transformations are as follows:
\begin{eqnarray}
	(j_{x},j_{y})\rightarrow(j_{y},-j_{x}),\quad 
	(E_{x},E_{y})\rightarrow(E_{y},-E_{x}). 
\end{eqnarray}
Under the time-reversal ($\mathcal{T}$) operation, their transformations are as follows:
\begin{eqnarray}
	(j_{x},j_{y})\rightarrow(-j_{x},-j_{y}), (E_{x},E_{y})\rightarrow(E_{x},E_{y}).
\end{eqnarray}
Therefore, under the $C_{4z}\mathcal{T}$ operation, the two equations in Eq.~(\ref{Ohm1}) 
become 
\begin{eqnarray}
	j_{y}=\sigma_{xy}E_{x},\quad j_{x}=\sigma_{yx}E_{y}.
\end{eqnarray}
By further using the antisymmetric property of the Hall conductivity tensor: $\sigma_{xy}=-\sigma_{yx}$, it is evident that $\sigma_{xy}$ vanishes identically. 

Next, we examine the component $\sigma_{zx}$. The corresponding equation for the current component and electric field component is 
\begin{eqnarray}
	j_{z}=\sigma_{zx}E_{x}.\label{zx1}
\end{eqnarray}
Similarly, under the $C_{4z}\mathcal{T}$ operation, the equation becomes
\begin{eqnarray}
	-j_{z}=\sigma_{zx}E_{y}.
\end{eqnarray}
By doing one more $C_{4z}\mathcal{T}$ operation, one obtains 
\begin{eqnarray}
	j_{z}=-\sigma_{zx}E_{x}.\label{zx2}
\end{eqnarray}
A combination of Eq.~(\ref{zx1}) and Eq.~(\ref{zx2}) immediately indicates $\sigma_{zx}=0$. The vanishing of  $\sigma_{zy}$
can similarly be determined. 

For the anomalous Hall effect to occur, the $C_{4z}\mathcal{T}$ symmetry must be broken. 
In this work, we introduce a Zeeman field described by $B_{z}\sigma_{z}$, which explicitly 
breaks the $C_{4z}\mathcal{T}$ symmetry. Such a Zeeman field 
can be generated by applying a magnetic field along the $z$ direction. Although the 
magnetic field also induces orbital effects that contribute to the ordinary Hall effect, 
we neglect these orbital effects, as a strong anomalous Hall effect can be achieved 
even with a weak Zeeman field in this system. We note that while the Zeeman field described by $B_{z}\sigma_{z}$ explicitly breaks 
the $C_{4z}\mathcal{T}$ symmetry, the Hamiltonian retains a $C_{2z}$ symmetry even in the presence of SOC. This preserved symmetry 
ensures that $\sigma_{zx}$ and $\sigma_{zy}$ remain vanishing. As a result, only $\sigma_{xy}$
needs to be investigated. 

In the main text, we have shown that, when $\lambda_{\rm so}=0$,  
a weak Zeeman field will change the CNLs to two nodal rings located at the two mirror-invariant 
planes at $k_{z}=0$ and $\pi$,  leading to the occurrence of three-dimensional 
quantized anomalous Hall effects when the Fermi surface coincides with these two nodal rings. The underlying 
physics can be understood by the layer Chern number. Specifically, from a dimension-reduction perspective, 
the two-dimensional layer Hamiltonian with a fixed $k_{z}$ describes a Chern insulator characterized by 
a Chern number of value  $\mathcal{C}=1$ or $-1$ (depending on the direction of the Zeeman field) 
when  $k_{z}\neq0$ and $\pi$. Accordingly, each layer contributes 
to a Hall conductivity $\sigma_{xy}=\frac{e^{2}}{h}$. By summing the contribution from all layers, one obtains 
\begin{eqnarray}
	\sigma_{xy}&=&\frac{e^{2}}{\hbar}\sum_{n=c/v}\int\frac{d^{3}k}{(2\pi)^{3}}\Omega_{z}^{(n)}f(E_{n})\nonumber\\
	&=&-\frac{e^{2}}{h}\int\frac{dk_{z}}{2\pi}\mathcal{C}(k_{z})=-\frac{e^{2}}{h}\frac{G_{z}}{2\pi}=\pm\frac{e^{2}}{h}.
\end{eqnarray} 
Here, $G_{z}$ represents the reciprocal lattice vector along the $z$ direction. 
If the lattice constant is restored, the expression of $\sigma_{xy}$ is $\pm\frac{e^{2}}{h}\frac{1}{a_{z}}$, where 
$a_{z}$ represents the lattice constant along the $z$ direction. This quantization requires that all $k_{z}$ planes 
except for $k_{z}=0$ and $k_{z}=\pi$ are gapped. In other words, it requires that the Fermi surface coincides with
the two nodal rings at the $k_{z}=0$ and $k_{z}=\pi$ planes. In our model, this corresponds to the conditions that $t=t_{z}=0$ and $\mu=0$
are to be satisfied.

\begin{figure}[t]
	\centering
	\subfigure{
		\includegraphics[scale=0.33]{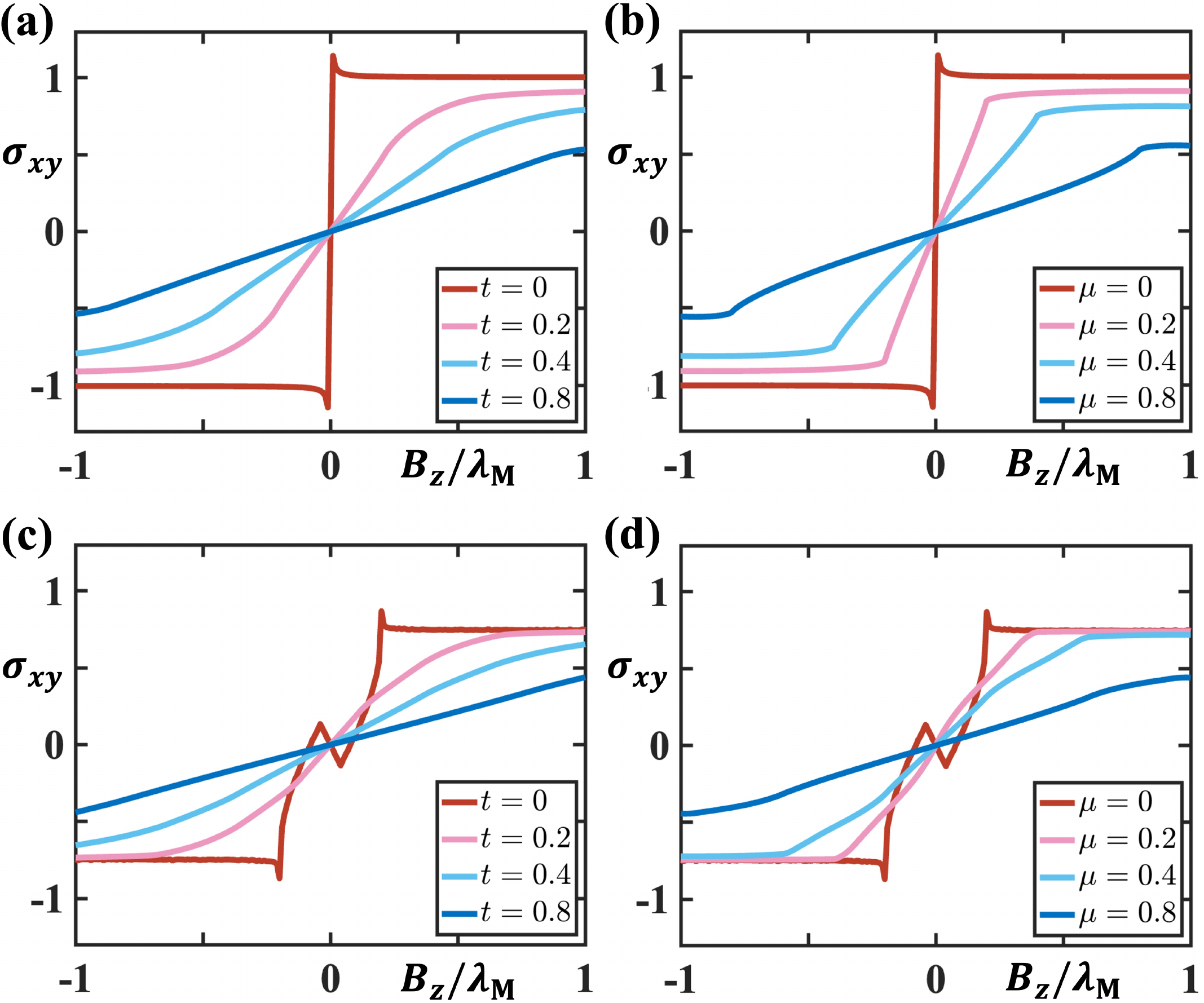}}
	\caption{{\bf The evolution of the Hall conductivity $\sigma_{xy}$ (in unit of $e^{2}/h$) with respect to $B_{z}$ for different hopping amplitudes [(a,c)] and chemical potential [(b,d)]}. We set $t=t_{z}$, $\mu=0$ in [(a,c)] and $t=t_{z}=0$ in [(b,d)]. The strength of spin-orbit coupling $\lambda_{\rm so}$ is set to $0$ in [(a-b)] and $0.2$  in [(c-d)], respectively. Common parameters are $\lambda_{\rm M}=1$ and $\eta=1$.}
	\label{SFig2}
\end{figure}

Here, we investigate the behavior of $\sigma_{xy}$ as $t$, $t_{z}$ and $\mu$ are varied across a range of values. 
When $t$ and $t_{z}$ becomes finite, or when $\mu$ is finite, the Fermi surface deviates from the two nodal rings. 
Our results, presented in Fig. \ref{SFig2}, demonstrates the breakdown of the Hall plateau when $t$ and $t_{z}$ becomes finite, 
as illustrated in Fig.~\ref{SFig2}(a).
Furthermore, the results indicate that in the weak field regime, the anomalous Hall effect 
diminishes as the hopping amplitudes increase. This occurs because  $t$ and $t_{z}$ 
makes the nodal lines dispersive. When $t$ and $t_{z}$ are large, the dispersion becomes pronounced, 
reducing the regions with divergent Berry curvature near the Fermi surface. 
Consequently, this leads to a weaker anomalous Hall effect. 

Intriguingly, if $t$ and $t_{z}$ remain zero, we observe that an approximate Hall plateau can emerge 
even for finite $\mu$, provided that the condition $B_{z}>\mu$ is satisfied. 
This phenomenon arises because the energy gaps in the $k_{z}$ planes (for $k_{z}\neq0$ and $k_{z}\neq\pi$)
increase as $B_{z}$. When $\mu$ lies within these energy gaps, the contributions from the $k_{z}$ planes 
become quantized. However, since this system is a semimetal, the energy gaps of the $k_{z}$ planes varies continuously
as a function of $k_{z}$. A larger $B_{z}$ enhances the $k_{z}$-dependence of this function, making it more sharply varying. 
Nevertheless, planes sufficiently close to $k_{z}=0$ and $k_{z}=\pi$ always exhibit nonzero and nonquantized contributions. Consequently, 
the plateau is approximate as long as $\mu$ becomes finite.

When SOC is introduced, we have previously demonstrated that, 
in addition to MKWNs at fixed positions, the interplay 
between SOC and the magnetic exchange field can generate 
additional Weyl nodes at generic positions within the Brillouin zone. When 
the Zeeman field is also included in this interplay, the dependence of the 
Weyl node distribution on the system's parameters is detailed in Table.~\ref{table: Weyl}.  

In the main text, we demonstrated that Hall plateaus also emerge when the band degeneracies evolve into Weyl nodes. The Hall plateau's  value is determined by the separation between these Weyl nodes, which are located in $k_{z}$ planes uniquely 
determined by the ratio of $\lambda_{\rm so}$ and $\lambda_{\rm M}$. The underlying mechanism 
can similarly be understood using the layer Chern numbers. Specifically, for $k_{z}$ 
planes without Weyl nodes, each plane is characterized by a nonzero Chern number. As a result, 
these planes contribute quantized values to the Hall conductivity when $\mu$
lies within the energy gap of the corresponding $k_{z}$  planes. Similarly, when 
$t$ and $t_{z}$ become finite, causing the Weyl nodes to separate in energy, 
we observe the breakdown of the Hall plateau, as illustrated in Fig.~\ref{SFig2}(c). 
Intriguingly, when $t$ and $t_{z}$
vanish, a Hall plateau can emerge even for finite $\mu$, provided that $B_{z}$ exceeds a 
$\mu$-dependent critical value, as illustrated in Fig.~\ref{SFig2}(d). Compared to the nodal-ring case, 
one can see that the Hall plateau is flatter in the Weyl-node case. This behavior arises because, 
for an untilted Weyl cone, the linear-order low-energy Hamiltonian, $\mathcal{H}(\bk)=\sum_{ij}v_{ij}q_{i}\sigma_{j}$,
where $\bq$ represents the momentum measured from the Weyl node and $v_{ij}$ denotes a velocity matrix, 
exhibits an emergent time-reversal symmetry that forces the contribution from the Weyl node to vanish.
For a large $B_{z}$, the energy window displaying a well-defined linear-dispersive spectrum becomes substantial. 
Consequently, a Hall plateau emerges when $\mu$ lies within this energy window.

In real materials, $t$ and $t_{z}$ are generally finite, making the observation 
of the predicted Hall plateau less feasible. Nevertheless, a strong anomalous Hall effect induced by a weak 
field remains an intriguing and experimentally observable phenomenon. As a final remark, 
since the direction of the Zeeman field strongly affects both {\zy the} symmetry and {\zy the} band structure, 
the anomalous Hall effect exhibits a pronounced angular dependence. {\zy Specifically, finite components $\sigma_{xz}$ and $\sigma_{yz}$ may emerge when an in-plane magnetic field is applied and the combined symmetries $C_{2x}\mathcal{T}$ and $C_{2y}\mathcal{T}$ are broken. However, these components are not quantized because the two-dimensional momentum planes at fixed $k_{x}$ or $k_{y}$ correspond to trivial insulators when their spectra are gapped.}
This angular dependence {\zy provides an additional means to probe} the band structure {\zy and the topology}, 
complementing techniques such as angle-resolved photoemission spectroscopy {\zy and magneto-optical Kerr effects}.

\begin{figure}[t]
	\centering
	\subfigure{
		\includegraphics[scale=0.45]{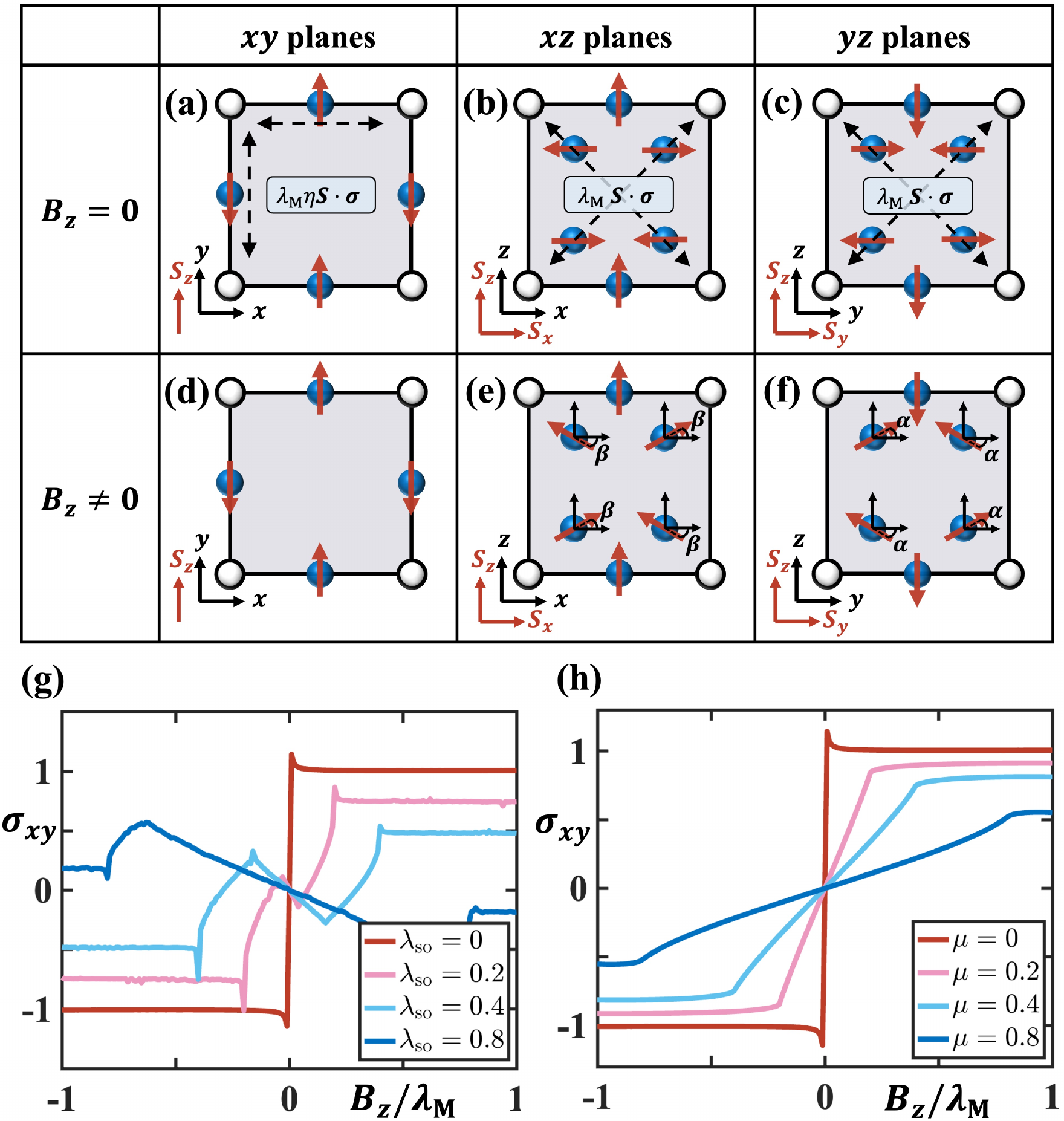}}
	\caption{{\bf Canting of the local magnetic moments and anomalous Hall effects.} (a-f) Schematic illustration of the orientations of the local magnetic moments in the absence [(a-c)] or presence [(d-f)] of a Zeeman field along the $z$ direction . The canting angle $\alpha$ ($\beta$) refers to the orientation deviation of the magnetic moments from $y$ ($x$) direction towards $z$ direction for magnetic atoms located in the $yz$ ($xz$) planes of the unit cell.
		(g-f) Hall conductivity tensor $\sigma_{xy}$ as a function of $B_{z}/\lambda_{\rm M}$ for the full Hamiltonian in Eq. (\ref{eq: pertur H}). Without loss of generality, we set $\delta_{1x}=\delta_{1y}=0.1B_{z}$. In [(g)], the chemical potential is fixed at $\mu=0$, while in [(h)], spin-orbit coupling is turned off ($\lambda_{\rm so}=0$). Other parameters are $t=t_{z}=0$ and $\lambda_{\rm M}=\eta=1$.}
	\label{SFig3}
\end{figure}

\section{IV. Zeeman field effects on the local magnetic moments}

In the main article, we {\zy approximate} the effect of an external magnetic field {\zy as introducing} a Zeeman term for itinerant electrons. This approximation is well 
justified for weak magnetic fields, as in this regime, while the weak Zeeman term 
strongly influences the Berry curvature distribution, it has minimal impact on the 
magnetic background formed by local magnetic moments. In this section, we examine the 
canting of local magnetic moments {\zy (or spin flop)} induced by {\zy the} Zeeman coupling{\zy. This canting is accompanied by a weak ferromagnetic component. In principle, such canting does not occur in weak magnetic field because the exchange interaction in antiferromagnets is typically strong, and a comparable Zeeman energy (corresponding to a magnetic field on the order of 1-10 T) is required to reorient the spins. Nevertheless, for completeness, we analyze this effect here} and demonstrate that, {\zy even for a weak spin canting, which itself already requires a large Zeeman field, the anomalous Hall effect discussed in Section III remains qualititively unchanged.} 

We begin with an energy scale analysis. Assume that the strength of magnetic field is less than 1T. The energy scale 
of the Zeeman term is of the order 0.1 meV (assuming the Landé factor $g=2$), which is much smaller 
than the typical energy scale of exchange interaction leading to high-temperature magnetic orders (e.g., 100 meV). 
This suggests that for materials whose magnetic order persists at relatively high temperatures 
(e.g., near room temperature), the influence of an external magnetic field on the magnetic order can generally be neglected. 
Nevertheless, how the local magnetic moment canting affects both the Hamiltonian and the AHE is an interesting 
question.  To explore this effect, we consider a Zeeman field along the $z$ direction, 
enabling a direct comparison between results with and without including this effect.

A Zeeman field along the $z$-direction breaks symmetries including $\mathcal{M}_{xy}$, $\mathcal{M}_{\bar{x}y}$ and $C_{4z}\mathcal{T}$ symmetries, while preserving $\mathcal{M}_{z}$, $\mathcal{M}_{x}\mathcal{T}$, and $\mathcal{M}_{y}\mathcal{T}$. The latter two symmetries indicate that the system remains invariant under the combination of a mirror reflection $\mathcal{M}_{x(y)}$ and time reversal $\mathcal{T}$. These conserved symmetries permit spin canting only along $z$, leading to a perturbation of the form: $\Delta \mathcal{H}(\bk)=\lambda_{\rm M}\sin k_{z}(\delta_{1x}\sin k_{x}+\delta_{1y}\sin k_{y})\sigma_{z}$, alongside a slight parameter renormalization of the spin-dependent hoppings in the $xz$ and $yz$ planes: $\lambda_{\rm M}\sin k_{z}(-\delta_{0x}\sin k_{x}\sigma_{x}+\delta_{0y}\sin k_{y}\sigma_{y})$. Physically, the parameters \{$\delta_{0x}$, $\delta_{1x}$\} and \{$\delta_{0y}$, $\delta_{1y}$\} should be functions of the canting angles $\beta$ and $\alpha$, respectively. We assume that the magnitude of each local magnetic moment is conserved, 
which imposes the normalization conditions: $\sqrt{\delta_{0x}^{2}+\delta_{1x}^{2}}=1$ and $\sqrt{\delta_{0y}^{2}+\delta_{1y}^{2}}=1$. 
A schematic of the orientations of these  moments with and without canting is shown in Figs.~\ref{SFig3} (a-f). Taking the perturbation into account, the full Hamiltonian becomes
\begin{eqnarray}
	\mathcal{H}(\bk)&=&-\left[t(\cos k_{x}+\cos k_{y})+t_{z}\cos k_{z}\right]+\lambda_{\rm so}(\sin k_{x}\sigma_{x}+\sin k_{y}\sigma_{y}+\sin k_{z}\sigma_{z})\nonumber\\
	&&+\lambda_{\rm M}\eta (\cos k_{x}-\cos k_{y})\sigma_{z}+\lambda_{\rm M}\sin k_{z}(-\delta_{0x}\sin k_{x}\sigma_{x}+\delta_{0y}\sin k_{y}\sigma_{y})\nonumber\\
	&&+\lambda_{\rm M}\sin k_{z}(\delta_{1x}\sin k_{x}+\delta_{1y}\sin k_{y})\sigma_{z}+B_{z}\sigma_{z}.
	\label{eq: pertur H}
\end{eqnarray}
Assuming a weak magnetic field, the canting angles $\alpha$, $\beta$ are expected to be small. Hence, without loss of generality, we assume that $\delta_{1x(y)}$ scales linearly with $B_{z}$. It is evident that the perturbation $\Delta\mathcal{H}(\bk)$ vanishes at momentum planes $k_{z}=\{0,\pi\}$, and does not introduce new band degeneracies. Thus, in the absence of {\zb SOC}, i.e., $\lambda_{\rm so}=0$, each $k_{z}$-slice with $k_{z}\neq\{0,\pi\}$ remains a Chern insulator with the Chern number $\mathcal{C}(k_{z})=-{\rm sgn}(B_{z})$, yielding a quantized Hall conductivity $\sigma_{xy}=-\frac{e^{2}}{h}{\rm sgn}(B_{z})$. With finite SOC, the quantized Hall response persists as long as the $k_{z}$-positions of the Weyl nodes remain insensitive to $B_{z}$. The perturbation only slightly shifts the Weyl nodes within the $k_{z}$-planes given by: $\sin k_{x}(\lambda_{\rm so}-\lambda_{\rm M}\delta_{0x}\sin k_{z})=\sin k_{y}(\lambda_{\rm so}+\lambda_{\rm M}\delta_{0y}\sin k_{z})=0$. Numerical calculations [Figs. \ref{SFig3}(g-h)] for this full Hamiltonian confirm that the Hall conductivity remains nearly unchanged (see the agreement between Fig.~\ref{SFig2}(b) 
and Fig.~\ref{SFig3}(h)). These results confirm that the canting effect to AHE can be safely neglected in weak magnetic fields.

\section{V. Connection to real-world materials}
In this section, we discuss potential material realizations of the CNLs and magnetic Kramers Weyl nodes predicted by our model. In principle, realizing CNLs in real systems requires three mutually orthogonal mirror planes---a condition naturally satisfied in antiferromagnets with noncollinear magnetic moments belonging to orthorhombic, tetragonal, or cubic point groups. Such magnetic systems provide promising platforms for realizing the physics described in our work.

A particularly compelling candidate is the recently reported spin-split antiferromagnet $\rm MnTe_{2}$, identified by Zhu {\it et al.}~\cite{Zhu2024}. Upon performing a coordinate transformation that redefines the in-plane momenta as $k_{1} = (k_{x} - k_{y})/\sqrt{2}$ and $k_{2} = (k_{x} + k_{y})/\sqrt{2}$, our model Hamiltonian maps onto a form that hosts band degeneracies consistent with those reported in $\rm MnTe_{2}$. The transformed Hamiltonian reads:
\begin{eqnarray}
	\mathcal{H}_{\rm CNL}(\bk)&=&-\lambda_{\rm M}\sin k_{z}\sin\frac{k_{1}+k_{2}}{\sqrt{2}}\sigma_{x}+\lambda_{\rm M}\sin k_{z}\sin\frac{k_{2}-k_{1}}{\sqrt{2}}\sigma_{y}+\lambda_{\rm M}\eta(\cos\frac{k_{1}+k_{2}}{\sqrt{2}}-\cos\frac{k_{2}-k_{1}}{\sqrt{2}})\sigma_{z}\nonumber\\
	&=&-\lambda_{\rm M}
	\left[\sin k_{z}\cos\frac{k_{1}}{\sqrt{2}}\sin\frac{k_{2}}{\sqrt{2}}(\sigma_{x}-\sigma_{y})+\sin k_{z}\sin\frac{k_{1}}{\sqrt{2}}\cos\frac{k_{2}}{\sqrt{2}}(\sigma_{x}+\sigma_{y})+2\eta\sin\frac{k_{1}}{\sqrt{2}}\sin\frac{k_{2}}{\sqrt{2}}\sigma_{z}\right]\nonumber\\
	&=&-\lambda_{\rm M}
	\left[\sqrt{2}\sin k_{z}\cos\frac{k_{1}}{\sqrt{2}}\sin\frac{k_{2}}{\sqrt{2}}\sigma_{-}+\sqrt{2}\sin k_{z}\sin\frac{k_{1}}{\sqrt{2}}\cos\frac{k_{2}}{\sqrt{2}}\sigma_{+}+2\eta\sin\frac{k_{1}}{\sqrt{2}}\sin\frac{k_{2}}{\sqrt{2}}\sigma_{z}\right]\nonumber\\
	&\equiv&-\lambda_{\rm M}
	\left[\sqrt{2}\sin k_{z}\cos\frac{k_{x}}{\sqrt{2}}\sin\frac{k_{y}}{\sqrt{2}}\sigma_{x}
	+\sqrt{2}\sin k_{z}\sin\frac{k_{x}}{\sqrt{2}}\cos\frac{k_{y}}{\sqrt{2}}\sigma_{y}
	+2\eta\sin\frac{k_{x}}{\sqrt{2}}\sin\frac{k_{y}}{\sqrt{2}}\sigma_{z}\right],
	\label{eq: transCNL}
\end{eqnarray}
where $\sigma_{\pm} = (\sigma_x \pm \sigma_y)/\sqrt{2}$. In the final line, we relabel $(k_{1}, k_{2}, \sigma_{-}, \sigma_{+})$ as $(k_x, k_y, \sigma_x, \sigma_y)$ for clarity and simpler comparison. In this rotated basis, the CNLs align with the high-symmetry paths $\Gamma$–$X_{1}$–$M$–$\Gamma$, consistent with the DFT-calculated band degeneracies in $\rm MnTe_{2}$ (see Fig.~1 of Ref.~\cite{Zhu2024}). Even more convincing is the agreement in spin textures: the signs of the bulk spin expectation values computed from our Hamiltonian Eq.~(\ref{eq: transCNL}) match those reported in DFT calculations~\cite{Zhu2024}. The above consistencies imply that the topological surface states, with unconventional spin textures, are expected on both $x$- and $y$-normal surfaces. Moreover, the $\bk \cdot \bp$ Hamiltonian near the $\Gamma$ point provided in Ref.~\cite{Zhu2024} ($h_{i}$ are scalar coefficients), 
\begin{eqnarray}
	H^{\Gamma}&=&\left[h_{1}+h_{3}(k_{x}^{2}+k_{y}^{2}+k_{z}^{2})\right]\sigma_{0}+h_{3}(k_{y}k_{z}\sigma_{x}+k_{z}{\zb k_{x}}\sigma_{y}+k_{x}k_{y}\sigma_{z})+O(\bk^{3}),
	\label{eq: MnTe2 Gamma}
\end{eqnarray}
matches exactly with the low-energy expansion of our model when $\eta = 1$:
\begin{eqnarray}
	\mathcal{H}_{\rm CNL}^{\Gamma}&=&-\lambda_{\rm M}(k_{y}k_{z}\sigma_{x}+k_{x}k_{z}\sigma_{y}+k_{x}k_{y}\sigma_{z}),
	\label{eq: CNL Gamma}
\end{eqnarray}
demonstrating a one-to-one correspondence at low energies.  Also, the absence of surface states on the $z$-normal surfaces, as reported in Ref.~\cite{Zhu2024}, is consistent with our theoretical prediction. Given these strong agreements, spin-ARPES measurements on the $x$- and $y$-normal surfaces of $\rm MnTe_{2}$ could be directly applied to detect the predicted surface states. 

{\zb Turn} to magnetic Kramers Weyl nodes. Since $\rm MnTe_{2}$ is centrosymmetric (belongs to point group $O_{h}$), spin-orbit coupled terms of the form $\lambda_{\rm so} \sum_{i=x,y,z} \sin k_{i} \sigma_{i}$ are forbidden, leading to the absence of magnetic Kramers Weyl nodes. However, symmetry-lowering perturbations---such as applied electric fields, uniaxial strain, interfacial engineering, or chemical doping---can induce such spin-orbit coupling while preserving $C_{4z}\mathcal{T}$ symmetry, thereby enabling magnetic Kramers Weyl nodes at symmetry-related TRIMs. Beyond $\rm MnTe_{2}$, chiral antiferromagnets such as $\rm Mn_{3}IrGe$ and $\rm YMnO_{3}$ have been predicted to host Weyl nodes at TRIMs by DFT calculations~\cite{Gao2023Heesch}. These materials also represent promising platforms for realizing the physics discussed here. Experimentally, magnetic Kramers Weyl nodes can manifest through long Fermi arcs (detectable via ARPES) and electromagnetic responses in linear and nonlinear regimes---including the anomalous Hall effect and circular photogalvanic effects. Notably, our model predicts quantized CPGE signals when the chemical potential is adjusted near the Weyl nodes.

Finally, we note that while exact CNL orthogonality arises in cubic, tetragonal, and orthorhombic systems, similar nodal degeneracies along high-symmetry lines may also arise in hexagonal and trigonal crystals, albeit with non-orthogonal configurations. Thus, the mechanisms uncovered in our study are applicable to a broad class of magnetic materials, spanning five of the seven crystal systems.

\end{widetext}

\end{document}